\documentclass[journal,twoside,web]{ieeecolor}
\usepackage{tmi}
\usepackage{cite}
\usepackage{amsmath,amssymb,amsfonts}
\usepackage{textcomp}
\usepackage{multirow}
\usepackage{algorithm}
\usepackage{algorithmic}
\usepackage{hyperref}
\usepackage{url}
\usepackage{graphicx}
\usepackage{subfigure}
\usepackage{amsmath,bm}
\newcommand{\jiaxing}[1]{{\color{black}#1}}
\newcommand{\minor}[1]{{\color{black}#1}}
\def\BibTeX{{\rm B\kern-.05em{\sc i\kern-.025em b}\kern-.08em
    T\kern-.1667em\lower.7ex\hbox{E}\kern-.125emX}}
\begin{document}
\title{Contrastive Graph Pooling for Explainable Classification of Brain Networks}
\author{Jiaxing Xu, Qingtian Bian, Xinhang Li, Aihu Zhang, Yiping Ke, Miao Qiao, \\ Wei Zhang, Wei Khang Jeremy Sim, and Balázs Gulyás \\ for the Alzheimer’s Disease Neuroimaging Initiative $^\dagger$
\thanks{Manuscript submitted 11 July 2023. This research/project is supported by the National Research Foundation, Singapore under its Industry Alignment Fund – Pre-positioning (IAF-PP) Funding Initiative, and the Ministry of Education, Singapore under its MOE Academic Research Fund Tier 2 (STEM RIE2025 Award MOE-T2EP20220-0006), and MBIE Catalyst: Strategic Fund NZ-Singapore Data Science Research Programme UOAX2001. Any opinions, findings and conclusions or recommendations expressed in this material are those of the author(s) and do not reflect the views of National Research Foundation, Singapore, and the Ministry of Education, Singapore.}
\thanks{Jiaxing Xu (corresponding author), Qingtian Bian, Aihu Zhang, and Yiping Ke are with the School of Computer Science and Engineering, Nanyang Technological University, Singapore, 639798 Singapore (e-mail:\url{JIAXING003@e.ntu.edu.sg}; \url{BIAN0027@e.ntu.edu.sg}; \url{ZHAN0547@e.ntu.edu.sg}; \url{ypke@ntu.edu.sg}).}
\thanks{Xinhang Li is with the Department of Computer Science and Technology, Tsinghua University
BEIJING, 100084 P.R.China (e-mail: \url{xh-li20@mails.tsinghua.edu.cn}).}
\thanks{Miao Qiao is with the School of Computer Science, the University of Auckland, Auckland 1142
New Zealand (e-mail: \url{miao.qiao@auckland.ac.nz}).}
\thanks{Wei Zhang, Wei Khang Jeremy Sim, and Balázs Gulyás are with the Lee
Kong Chian School of Medicine, Nanyang Technological University,
Singapore, 636921 Singapore (e-mail: \url{wilson.zhangwei@ntu.edu.sg}; \url{SIMW0035@e.ntu.edu.sg}; \url{balazs.gulyas@ntu.edu.sg}).}
\thanks{$^\dagger$Part of the data used in preparation of this article were obtained from the Alzheimer’s Disease Neuroimaging Initiative (ADNI) database (adni.loni.usc.edu). As such, the investigators within the ADNI contributed to the design and implementation of ADNI and/or provided data but did not participate in the analysis or writing of this report. A complete listing of ADNI investigators can be found at: \url{http://adni.loni.usc.edu/ wp-content/uploads/how_to_apply/ADNI_Acknowledgement_List.pdf}}}

\maketitle

\begin{abstract}
Functional magnetic resonance imaging (fMRI) is a commonly used technique to measure neural activation. Its application has been particularly important in identifying underlying neurodegenerative conditions such as Parkinson's, Alzheimer's, and Autism. Recent analysis of fMRI data models the brain as a graph and extracts features by graph neural networks (GNNs). However, the unique characteristics of fMRI data require a special design of GNN. Tailoring GNN to generate effective and domain-explainable features remains challenging. In this paper, we propose a contrastive dual-attention block and a differentiable graph pooling method called \textit{ContrastPool} to better utilize GNN for brain networks, meeting fMRI-specific requirements. We apply our method to 5 resting-state fMRI brain network datasets of 3 diseases and demonstrate its superiority over state-of-the-art baselines. Our case study confirms that the patterns extracted by our method match the domain knowledge in neuroscience literature, and disclose direct and interesting insights. Our contributions underscore the potential of ContrastPool for advancing the understanding of brain networks and neurodegenerative conditions. \minor{The source code is available at \url{https://github.com/AngusMonroe/ContrastPool}.} 
\end{abstract}

\begin{IEEEkeywords}
Brain Network, Deep Learning for Neuroimaging, fMRI Biomarker, Graph Classification, Graph Neural Network.
\end{IEEEkeywords}

\section{Introduction}
\label{sec:introduction}
\IEEEPARstart{F}{unctional} magnetic resonance imaging (fMRI) measures neural activation and has been widely used in analyzing functional connectivities of the human brain \cite{worsley2002general}. Particularly, fMRI has been instrumental in identifying underlying neurodegenerative conditions \cite{poldrack2009decoding}, e.g., Alzheimer’s, Parkinson’s, and Autism.

A line of research in analyzing fMRI data models brains as graphs, i.e., defining brain regions of interest (ROIs) as nodes and the functional connectivities/interactions between ROIs as edges \cite{xu2023data}. Graph \jiaxing{analytic} methods are then applied to such brain networks to understand and analyze the elements and interactions of neurological systems from a network perspective. While previous research has showcased the remarkable potential of network-based approaches \cite{kawahara2017brainnetcnn,lanciano2020explainable} in deciphering the complexities of the brain, graph \jiaxing{analysis} for brain networks is still an emerging field in its early stages of development.

Recently graph neural networks (GNNs) have attracted increasing interest in graph classification and thus have been applied to graph-structured data in a wide range of applications, including social networks \cite{kipf2016semi}, molecular data \cite{gilmer2017neural,xu2024union} and recommendation systems \cite{wu2022graph,bian2023cpmr}. Compared with conventional machine learning models that handle vector-based data, GNNs are able to engage graph topological information through message passing. In light of the promising performance of GNNs on other applications, several studies \cite{ktena2017distance,li2020pooling,li2021braingnn}  have applied GNNs to brain network analysis. However, directly applying general-purpose GNNs to brain networks could be ineffective on fMRI data due to its unique data characteristics \cite{yan2019groupinn}\jiaxing{:}
(1) \textbf{Low signal-to-noise ratio (SNR).} fMRI is a type of high dimensional data where non-neural noise derived from cardiac/respiratory processes or scanner instability could cause large variations within a single subject and across different subjects. 
(2) \textbf{Node alignment.} Different from other graph-based datasets, each node of a brain network represents an ROI under a brain parcellation scheme. If the same scheme is applied to all subjects, each resulting network will have the same number of nodes and the nodes are aligned across different subjects. 
(3) \textbf{Limited data scale.} Due to the limited availability of the fMRI data, brain network datasets contain a relatively small number of subjects, which can cause overfitting \jiaxing{on} GNNs.

To better utilize GNNs for brain networks and meet the need \jiaxing{for} fMRI characteristics, we propose a contrastive graph pooling method (\textit{ContrastPool}) with a contrastive dual-attention to extract group-specific information. \jiaxing{ In contrastPool, the ROI-wise attention is introduced to identify the most representative brain regions in each class of subjects for the given task (e.g., the Alzheimer's disease classification). This design aims to reduce the noise caused by the fMRI acquisition process by directing the model to focus on just a few ROIs rather than the entire set, which alleviates the issue of low SNR. In addition, ContrastPool devises a subject-wise attention that leverages the node alignment in brain networks. It aligns different subjects in the same class by ROI and assigns higher attention scores to subjects that are more indicative of the class. By performing aggregation within groups and contrast across groups, ContrastPool generates a contrast graph and uses it to guide the message passing in brain network representation learning. The group knowledge can weaken the sensitivity to specific subjects and thus mitigates the overfitting issue. The contrast operation further identifies the discrepancy across groups and strengthens the dual attention to focus on more discriminative ROIs and subjects.} 
\jiaxing{Different from graph contrastive learning methods \cite{you2020graph,yin2022autogcl,chen2023attribute} which construct positive and negative pairs by data augmentation, we make contrast across groups. The rationale behind this design is to identify the discrepant features discovered from different groups.  Therefore, we use a contrast graph to extract the most discriminative task-related features from the training set and use it as prior knowledge to guide brain network representation learning.} Our key contributions are:
\begin{itemize}
\item We propose a contrastive dual-attention block, which adaptively assigns a weight to each ROI of each subject, \jiaxing{performs adaptive aggregation over subjects in each group, and makes contrast across different groups to obtain a contrast graph. Guided by the contrast graph, we introduce a differentiable graph pooling method called \textit{ContrastPool} to generate brain network representations that are effective to the task of disease classification.}
\item We apply our ContrastPool to 5 resting-state fMRI brain network datasets spanning over 3 diseases. The results justify the superiority of ContrastPool over the state-of-the-art baselines.
\item The case study confirms the interestingness, simplicity and high explainability of the patterns extracted by our method, which match the domain knowledge in neuroscience literature. 
\end{itemize}

\section{Related Work}

\subsection{General-purposed GNNs}

\noindent
\jiaxing{\textbf{GNNs with Attention Mechanism.} The integration of attention mechanisms in GNNs \cite{velivckovic2017graph,brody2021attentive} involves incorporating attention-based mechanisms to selectively weigh the importance of different neighbors during the information aggregation process. This attention mechanism enhances the modeling of relationships between nodes in a graph, allowing the model to focus on more relevant neighbors. The relevancy of a neighbor is decided by the dynamic and automated assignment of the attention weights. Such an assignment is based on the content and features of neighboring nodes, which enables the GNN to capture complex and non-linear dependencies in graph-structured data, and thus leads to enhanced performances in various tasks \cite{choi2017gram,ma2017dipole}. GAT \cite{velivckovic2017graph} first introduces the attention mechanism to GNNs by utilizing the similarity of two endpoints to control the weight of the edge for message passing. AGNN \cite{thekumparampil2018attention} removes all intermediate fully connected layers of graph convolutional network and replaces the propagation layer with an attention mechanism that respects the graph structure. GAM \cite{lee2018graph} proposes structural attention to enhance graph structure learning for graph classification. Different from these works that introduce attentions to neighbors/graph structures and use the attentions directly in message passing, our work imposes attentions on ROIs and subjects in brain network analysis. Our dual attentions are coupled with group aggregation and contrast to produce a contrast graph, which is then used to guide the message passing. In this sense, our attention mechanism is not applied directly to the message-passing process of GNNs.}

\noindent
\jiaxing{\textbf{Graph Contrastive Learning.} In typical graph contrastive learning, positive and negative samples are generated by data augmentation on the graph structure or node features \cite{you2020graph}. SUGAR \cite{sun2021sugar} generates subgraphs and uses these subgraphs for reconstruction. AutoGCL \cite{yin2022autogcl} employs a set of learnable graph view generators orchestrated by an auto augmentation strategy, where every graph view generator learns a probability distribution of graphs conditioned by the input. ASP \cite{chen2023attribute} proposes a novel attribute and structure-preserving graph contrastive learning framework. The objective of these methods is to maximize the similarities of positive pairs and minimize the similarities of negative pairs. On the contrary, our contrastive dual-attention aims to highlight the discrepancy across different groups, which is based on ground-truth labels instead of generated samples.}

\noindent
\jiaxing{\textbf{Graph Pooling.} Graph pooling is a technique in graph neural networks that downsamples graphs by aggregating information from groups of nodes, reducing graph size while preserving essential structural and contextual features for efficient and effective learning. Existing graph pooling methods often select nodes or generate super-nodes by using attention mechanisms \cite{lee2019self}, maximizing mutual information \cite{nouranizadeh2021maximum} or encoding graph structures \cite{zhang2019hierarchical}. However, these methods consider the feature and structural information of each graph individually without considering the similarity within groups and the discrepancy across groups. }

\subsection{Models for Brain Networks}

In recent years, several GNN-based methods have been proposed for brain networks. \cite{ktena2017distance} leverages graph convolutional networks (GCNs) for learning similarities between each pair of graphs (subjects). \cite{kawahara2017brainnetcnn} proposes edge-to-edge, edge-to-node and node-to-graph convolutional filters to leverage the topological information of brain networks in the neural network. \cite{li2019graph} puts forward a two-stage pipeline to discover ASD brain biomarkers from task-fMRI using GNNs. \cite{li2021braingnn} proposes an ROI-selection pooling to highlight salient ROIs for each individual. 
\jiaxing{MG2G \cite{xu2021graph} adopts a two-stage approach that first learns latent distribution-based node representations by an unsupervised stochastic graph embedding model and then employs the learnt representations to train a classifier. The distribution-based nature of node representations enables the identification of significant ROIs with AD-related effects. These studies neglect the characteristics of fMRI data elaborated earlier in their model design. These works neglect the three characteristics of fMRI data elaborated in Section \ref{sec:introduction}.} Besides, \cite{li2020pooling} proposes a graph pooling method with group-level regularization to guarantee group-level consistency. \cite{yan2019groupinn} jointly learns the node grouping and extracts \jiaxing{the} graph features. 
These two methods only take group information into account \jiaxing{on} graph level without utilizing node alignment. \cite{lanciano2020explainable} extracts a dense contrast subgraph to filter useful information for prediction. However, their feature extraction and subject classification treat all ROIs and all subjects equally,  which could be vulnerable to noisy data. \cite{zhang2022classification} incorporates local ROI-GNN and global subject-GNN guided by non-imaging data, such as gender, age, and acquisition site. The local ROI-GNN does not take the node alignment of brain networks into account. \minor{Ding et al. \cite{ding2023parkinson} also integrate embeddings from separate graph views of images and non-imaging data by a co-attention module, along with a contrastive loss-based fusion method to improve cross-view fusion learning. Their contrast is applied to the population graphs of two modalities rather than groups, to facilitate effective fusion of information from different data modalities.} Moreover, \cite{kim2021learning} and \cite{yu2022learning} propose a graph generator to transform the raw blood-oxygen-level-dependent (BOLD) time series into task-aware brain connectivities. They model subjects as dynamic graphs, which require all subjects to have the same scan length and preferably from the same site. In contrast, our work constructs static brain networks as we operate on datasets collected from multi-sites with various acquisition lengths. A Transformer-based method \cite{kan2022brain} has been applied to brain networks to learn pairwise connection strengths among ROIs across individuals. It neglects the group information of subjects in its methodology design.

\section{Methodology}

In this section, we first introduce the construction of brain networks and commonly-used GNN schemes for graph classification. We then elaborate the design of our proposed ContrastPool. \jiaxing{Notation-wise, we use calligraphic letters (e.g., $\mathcal{G}$) to denote sets, bold capital letters (e.g., $\boldsymbol{A}$) matrices, and bold lowercase letters (e.g., $\boldsymbol{z}$) vectors.} Subscripts and superscripts are used to distinguish between different variables or parameters, while lowercase letters denote scalars. We use $\boldsymbol{A}(i,:)$ and $\boldsymbol{A}(:,j)$ to denote the $i$-th row and $j$-th column of a matrix $\boldsymbol{A}$, respectively. This notation also extends to a \jiaxing{3D} matrix. Table \ref{tab:notation} summarizes the notations used throughout the paper.

\begin{table}[h]
\renewcommand\arraystretch{1.26} 
\centering
\caption{Notation Table}
\begin{tabular}{cc}
\hline
Notation                         & Description                       \\ \hline
$\boldsymbol{M}$                 & Connectivity matrix of a subject                 \\
$G$                              & An input graph/brain network                             \\
$\boldsymbol{A}$                 & Adjacency matrix of $G$                 \\
$\boldsymbol{X}$                 & Node feature matrix of $G$                     \\
$\mathcal{V}_G$                  & Node set of $G$             \\
$v, u$                              & A node in $G$                              \\
$\boldsymbol{H}$                 & Node representations              \\
$\boldsymbol{H}_v$                & Node representation of $v$         \\
$m$                              & Number of nodes in $G$          \\
$\mathcal{D}$                    & Input dataset                         \\
$\mathcal{G}$                    & Input graph set                        \\
$\mathcal{Y}$                    & Input label set                        \\
$y_G$                            & Label of $G$                \\
$l$                              & The layer index in GNN    \\
$d$        & Dimensionality of node representation $\boldsymbol{H}_v$\\
$\mathcal{G}^{TC}, \mathcal{G}^{ASD}$  & Graph set of TC/ASD group    \\
$\mathcal{A}^{TC}, \mathcal{A}^{ASD}$  & Adjacency matrix set of TC/ASD group    \\
$\mathcal{X}^{TC}, \mathcal{X}^{ASD}$  & Feature matrix set of TC/ASD group    \\
$G_{sum}^{TC}, G_{sum}^{ASD}$ & Summary graph of TC/ASD group\\
$\boldsymbol{A}^{TC}_{sum}, \boldsymbol{A}^{ASD}_{sum}$ & Summary adjacency matrix of TC/ASD group    \\
$\boldsymbol{X}^{TC}_{sum}, \boldsymbol{X}^{ASD}_{sum}$ & Summary feature matrix of TC/ASD group    \\
$G_{contrast}$ & Contrast graph \\
$\boldsymbol{A}_{contrast}$         & Adjacency matrix of $G_{contrast}$      \\
$\boldsymbol{H}_{contrast}$         & Node representations of $G_{contrast}$      \\
$\boldsymbol{W}_Q, \boldsymbol{W}_K, \boldsymbol{W}_V$                 & Parameter matrix   \\
$i, j$                & Index for matrix dimensions            \\
$n^{TC}$              & Number of subjects in TC group \\
$\boldsymbol{T}_{ROI}^{TC}$         & Output \jiaxing{3D} matrix of $\operatorname{Attn}_{ROI}(\cdot)$ for TC group \\
$\boldsymbol{T}_{subject}^{TC}$         & Output \jiaxing{3D} matrix of $\operatorname{Attn}_{subject}(\cdot)$ for TC group \\
$\boldsymbol{S}$                 & Cluster assignment matrix            \\
$\boldsymbol{Z}$                 & Embedded node feature matrix             \\ 
$\boldsymbol{A}_{contrast}(i,:)$    & The $i$-th row of $\boldsymbol{A}_{contrast}$ \\\hline
\end{tabular}
\label{tab:notation}
\end{table}

\subsection{Preliminaries}

\noindent
\textbf{Brain Network Construction.}  We construct five brain network datasets from five neuro-imaging data sources, which are Taowu, PPMI and Neurocon \cite{badea2017exploring} for Parkinson's disease, ADNI \cite{dadi2019benchmarking} for Alzheimer's disease, and ABIDE \cite{craddock2013neuro} for Autism. 

\noindent
\emph{[TaoWu and Neurocon]} The TaoWu and Neurocon datasets, released by ICI \cite{badea2017exploring}, are among the earliest image datasets made available for Parkinson's research. These datasets comprise age-matched subjects collected from a single machine and site, encompassing both normal control individuals and patients diagnosed with PD.

\noindent
\emph{[PPMI]} The Parkinson's Progression Markers Initiative (PPMI) is a comprehensive study aiming to identify biological markers associated with Parkinson's risk, onset, and progression. PPMI comprises multimodal and multi-site MRI images. The dataset consists of subjects from 4 distinct classes: normal control (NC), scans without evidence of dopaminergic deficit (SWEDD), prodromal, and Parkinson's disease (PD). This ongoing study plays a crucial role in advancing our understanding of Parkinson's disease. 

\noindent
\emph{[ADNI]} The ADNI raw images used in this paper were obtained from the Alzheimer’s Disease Neuroimaging Initiative (ADNI) database (\url{adni.loni.usc.edu}). The ADNI was launched in 2003 as a public-private partnership, led by Principal Investigator Michael W. Weiner, MD. The primary goal of ADNI has been to test whether serial magnetic resonance imaging (MRI), positron emission tomography (PET), other biological markers, and clinical and neuropsychological assessment can be combined to measure the progression of mild cognitive impairment (MCI) and early Alzheimer’s disease (AD). For up-to-date information, see \url{www.adni-info.org}. We included subjects from 6 different stages of AD, from cognitive normal (CN), significant memory concern (SMC), mild cognitive impairment (MCI), early MCI (EMCI), late MCI (LMCI) to Alzheimer's disease (AD). 

\noindent
\emph{[ABIDE]} The Autism Brain Imaging Data Exchange (ABIDE) initiative supports the research on Autism Spectrum Disorder (ASD) by aggregating functional brain imaging data from laboratories worldwide. ASD is characterized by stereotyped behaviors, including irritability, hyperactivity, depression, and anxiety. Subjects in the dataset are classified into two groups: typical controls (TC) and individuals diagnosed with ASD.

\begin{table}[h]
\centering
\caption{Statistics of Brain Network Datasets.}
\begin{tabular}{lcccc}
\hline
Dataset & Condition & Graph\# & Avg BOLD length & Class\# \\ \hline
Taowu            & Parkinson          & 40      & 239             & 2       \\
PPMI             & Parkinson          & 209     & 198             & 4       \\
Neurocon         & Parkinson          & 41      & 137             & 2       \\
ADNI             & Alzheimer          & 1326    & 344             & 6       \\
ABIDE            & Autism             & 989     & 201             & 2       \\ \hline
\end{tabular}
\label{tab:dataset_statistic}
\end{table}

We followed \cite{xu2023data} in resting-state fMRI data preprocessing and brain network construction. Specifically, raw MRI images are collected from the five data sources. Each fMRI image is accompanied with a structural T1-weighted (T1w) image acquired from the same subject in the same scan session. 
The fMRI images are 
then preprocessed using fMRIPrep \cite{esteban2019fmriprep}, with T1w used to extract the brain mask and perform image alignment and normalization of BOLD signals. The preprocessing involves a number of pipeline steps, including motion correction, realigning, field unwarping, normalization, bias field correction, and brain extraction. Schaefer atlas \cite{schaefer2018local} is applied to the preprocessed images to parcellate the brain into 100 \jiaxing{ROIs}. A brain network is constructed for each subject, which is represented by a connectivity matrix $\boldsymbol{M}$. The nodes in $\boldsymbol{M}$ are ROIs and the edges are the Pearson's correlation between the region-averaged BOLD signals from pairs of ROIs. Essentially, $\boldsymbol{M}$ captures functional relationships between different ROIs. Statistics of the brain network datasets are summarized in Table \ref{tab:dataset_statistic}. \jiaxing{The class/group distribution of the datasets is provided in Table \ref{tab:dataset_class_distribution}.}

\begin{table}[h]
    \caption{\jiaxing{Class Distribution of Brain Network Datasets.}}
    \centering
    \begin{tabular}{ccc}
    \hline
       Dataset                 & Class   & \# Subjects \\
    \hline
        \multirow{2}{*}{ABIDE}    & Control    & 534 \\
                                  & ASD        & 455 \\
    \hline
        \multirow{6}{*}{ADNI}     & CN    & 819 \\
                                  & SMC        & 72  \\
                                  & LMCI       & 102 \\
                                  & EMCI       & 89  \\
                                  & MCI        & 179 \\
                                  & AD         & 65  \\
    \hline
        \multirow{4}{*}{PPMI}     & Control    & 15  \\
                                  & SWEDD      & 14  \\
                                  & Prodromal  & 67  \\
                                  & PD         & 113 \\
    \hline
        \multirow{2}{*}{TaoWu}    & Control    & 20 \\
                                  & PD         & 20 \\
    \hline
        \multirow{2}{*}{Neurocon} & Control    & 15 \\
                                  & PD         & 26 \\
    \hline
    \end{tabular}
    \label{tab:dataset_class_distribution}
\end{table}

\begin{figure*}[h]
\begin{center}
\includegraphics[width=.9\linewidth]{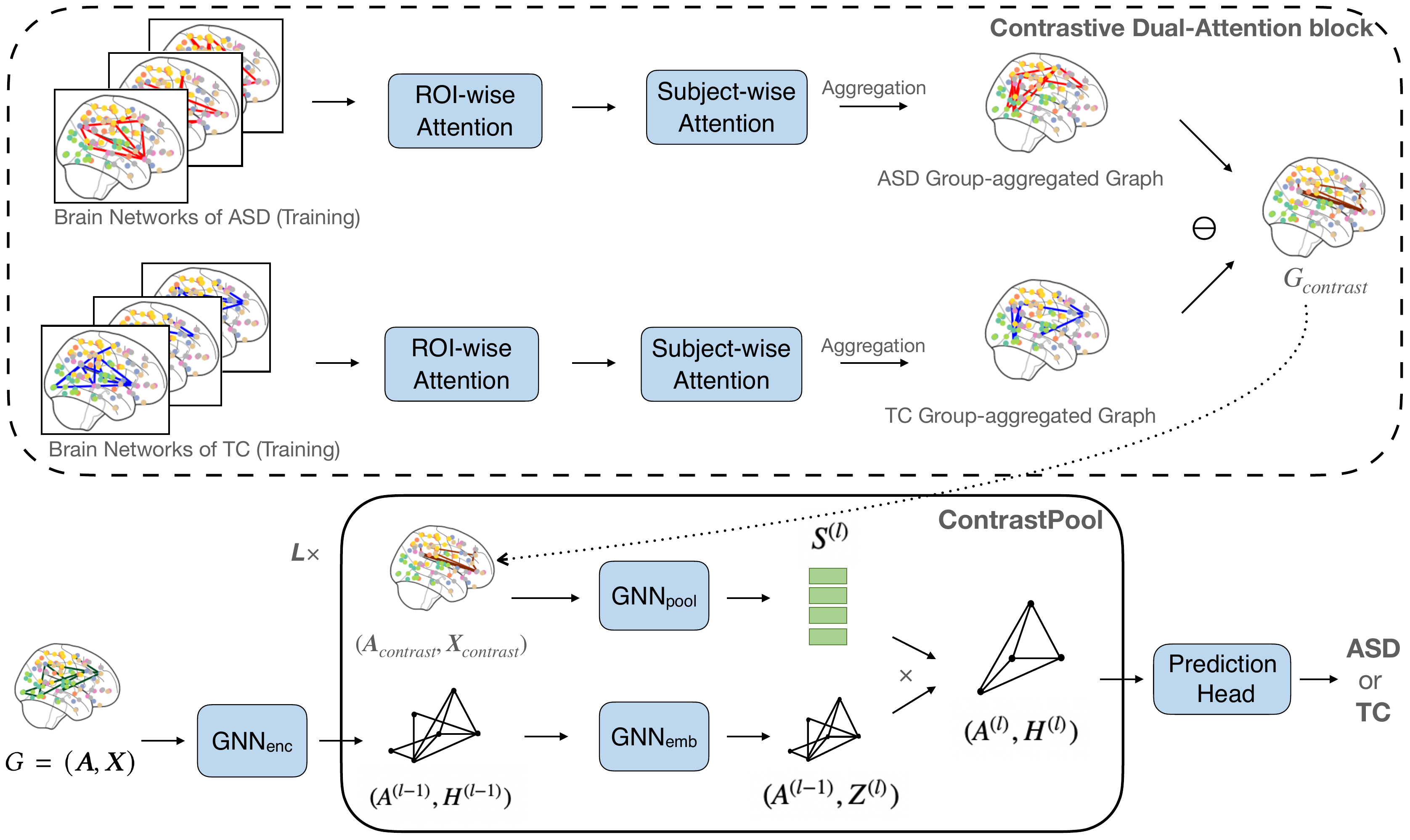}
\end{center}
\caption{The architecture of ContrastPool, using Autism as an example.}
\label{fig:framework}
\end{figure*}

\noindent
\textbf{Brain Network Classification.} Generally, graph classification aims to predict certain properties (i.e., class labels) of given graphs. Given a dataset of labeled graphs $\mathcal{D} = (\mathcal{G}, \mathcal{Y}) = \{(G, y_G)\}$, where $y_G$ is the class label of a graph $G \in \mathcal{G}$, the problem of graph classification is to learn a predictive function $f$: $\mathcal{G} \rightarrow \mathcal{Y}$, which maps input graphs to their labels. Take brain networks constructed from the Autism dataset as an example. The class label set contains two labels: TC and ASD\jiaxing{.}
The objective of brain network classification is to learn a predictive function $f$ from a set of brain networks, expecting that $f$ also works well on unseen brain networks.

\noindent
\textbf{Graph Neural Networks.} Compared with conventional vector-based machine learning models, GNNs engage graph topological information in graph representation learning. Consider an input graph $G = (\boldsymbol{A}, \boldsymbol{X})$, where $\boldsymbol{A}$ is the adjacency matrix, which encodes connectivity information, and $\boldsymbol{X}$ is the feature matrix containing attribute information for each node. The node set of $G$ is denoted as $\mathcal{V}_G$ and $|\mathcal{V}_G| = m$.
The $l$-th layer of a GNN in the message-passing scheme \cite{xu2018powerful} can be written as:
\begin{equation}
\label{eq:mpnn}
\boldsymbol{H}_{v}^{(l)}\!=\!\operatorname{AGG}^{(l-1)}\!\left(\boldsymbol{H}_{v}^{(l-1)}, \operatorname{MSG}^{(l-1)}\!\left(\left\{\!\boldsymbol{H}_{u}^{(l-1)}\!\right\}_{\forall u \in \mathcal{N}(v)}\!\right)\!\right).
\end{equation}

Herein, $\boldsymbol{H}^{(l)} \in \mathbb{R}^{m \times d}$ denotes the $l$-th layer node representation, where each node is represented by a $d$\jiaxing{-}dimensional vector. $\operatorname{AGG}(\cdot)$ and $\operatorname{MSG}(\cdot)$ are arbitrary differentiable aggregate and message functions (e.g., a multilayer perceptron (MLP) can be used as $\operatorname{AGG}(\cdot)$ and a summation function as $\operatorname{MSG}(\cdot)$). $\mathcal{N}(v)$ represents the neighbor node set of node $v \in \mathcal{V}_G$, and $\boldsymbol{H}^{(0)}_v = \boldsymbol{X}_v$. 
\jiaxing{The updated representations then pass through a sum/mean pooling and are fed to a linear layer for classification.} In brain network classification, we use the connectivity matrix $\boldsymbol{M}$ as both the adjacency matrix $\boldsymbol{A}$ and feature matrix $\boldsymbol{X}$. Despite prevalent, GNNs implementing Eq. (\ref{eq:mpnn}) only propagate information across the edges of each individual brain network, and thus neglect the group-based information and node alignment. This standard message passing scheme is also vulnerable to noise as it treats all ROIs and all subjects equally important in its representation learning process. Therefore, the goal of this work is to design an end-to-end framework that addresses these limitations of GNNs when applied to brain networks. 

\subsection{Contrastive Graph Pooling with Dual-Attention}

We propose ContrastPool, which addresses the above challenges by aggregating subjects over different classes using dual-attention. The idea of ContrastPool is based on an important observation: the relatedness of different ROIs for different diseases varies, and the extent to which a subject exhibits typical characteristics of a disease also varies. Therefore, we design ContrastPool in a way that the graphs in the same group (e.g., ASD group) are summarized by assigning  different weights for ROIs and subjects. These weights are not assigned manually  but learnt automatically from training data to optimize the classification performance.
The architecture of ContrastPool is shown in Fig. \ref{fig:framework}, with Autism as an example. In the training stage, all training graphs are categorized by their groups and \jiaxing{pass} through the Contrastive Dual-Attention (CDA) block to generate the contrast graph (the upper block of Fig. \ref{fig:framework}). This contrast graph is subsequently encoded into an assignment matrix, which is combined with each input graph for hierarchical graph pooling (the lower block of Fig. \ref{fig:framework}). In the validation and test stages, each validation/test graph \jiaxing{passes} through the lower block and \jiaxing{is fused} with the contrast graph (generated by the training phase) to obtain its graph representation. This representation is then fed into the Prediction Head to produce the final prediction of the graph (either ASD or TC in the example).

In the following, we first introduce the Contrastive Dual-Attention block, which is used to generate a contrast graph. We then describe how the ContrastPool module produces node and graph representations based on the contrast graph. Finally, we discuss our design of the loss function.

\noindent
\textbf{Contrastive Dual-Attention (CDA) Block.} This block aims to generate a contrast graph that best characterizes the differences in brain networks \jiaxing{between} two groups of subjects. In order to achieve this, it first summarizes all the graphs within the same group into a summary graph. It then computes the difference of the summary graphs from two different groups. We design the computation \jiaxing{between} the summary graph as a learnable dual-attention process such that the most discriminative ROIs and the most representative subjects can be automatically highlighted in the summary graph formation. 

Consider two groups of graphs in the training set, $\mathcal{G}^{TC} = (\mathcal{A}^{TC}, \mathcal{X}^{TC})$ and $\mathcal{G}^{ASD} = (\mathcal{A}^{ASD}, \mathcal{X}^{ASD})$, bearing TC and ASD groups, respectively. CDA first computes the summary graphs $G^{TC}_{sum}$ and $G^{ASD}_{sum}$ of the two groups via two subject aggregation functions $\operatorname{SA}_A(\cdot)$ and $\operatorname{SA}_X(\cdot)$ performed respectively on the adjacency matrix set $\mathcal{A}^{TC}$, $\mathcal{A}^{ASD}$ and the feature matrix set $\mathcal{X}^{TC}$, $\mathcal{X}^{ASD}$:

\begin{equation}
\label{eq:summary_g_tc}
G^{TC}_{sum} = (\boldsymbol{A}_{sum}^{TC}, \boldsymbol{X}_{sum}^{TC}) = (\operatorname{SA}_A(\mathcal{A}^{TC}), \operatorname{SA}_X(\mathcal{X}^{TC})),
\end{equation}

\begin{equation}
\label{eq:summary_g_asd}
G^{ASD}_{sum} = (\boldsymbol{A}_{sum}^{ASD}, \boldsymbol{X}_{sum}^{ASD}) = (\operatorname{SA}_A(\mathcal{A}^{ASD}), \operatorname{SA}_X(\mathcal{X}^{ASD})).
\end{equation}

The contrast graph $G_{contrast} = (\boldsymbol{A}_{contrast}, \boldsymbol{H}_{contrast})$ is then obtained by:


\begin{equation}
\label{eq:contrast_a}
\boldsymbol{A}_{contrast} = \boldsymbol{A}_{sum}^{TC} \ominus \boldsymbol{A}_{sum}^{ASD},
\end{equation}

\begin{equation}
\label{eq:contrast_h}
\boldsymbol{H}_{contrast} = \boldsymbol{X}_{sum}^{TC} \ominus \boldsymbol{X}_{sum}^{ASD},
\end{equation}

\noindent
where $\ominus$ is a binary function that computes the element-wise absolute differences of two matrices. 

To enable the automatic learning of the weights on the ROIs and subjects, respectively, we design a dual-attention function as $\operatorname{SA}(\cdot)$. In general, a self-attention function \cite{vaswani2017attention} for a given matrix $\boldsymbol{X} \in \mathbb{R}^{k \times m}$ can be written as:
\begin{equation}
\label{eq:attn}
\operatorname{Attn}(\boldsymbol{X}) = \operatorname{norm} \left(\boldsymbol{X} + \phi\left(\frac{\boldsymbol{Q} \boldsymbol{K}^\mathsf{T}}{\sqrt{k}}\right) \boldsymbol{V}\right),
\end{equation}

\begin{equation}
\label{eq:qkv}
\boldsymbol{Q}=\boldsymbol{X}\boldsymbol{W}_Q, \boldsymbol{K}=\boldsymbol{X}\boldsymbol{W}_K, \boldsymbol{V}=\boldsymbol{X}\boldsymbol{W}_V,
\end{equation}

\noindent
where $\boldsymbol{W}_Q, \boldsymbol{W}_K, \boldsymbol{W}_V \in \mathbb{R}^{m \times m}$ are parameter matrices, $\phi(\boldsymbol{z})_i=\operatorname{softmax}(\boldsymbol{z})_i=\frac{e^{\boldsymbol{z}_i}}{\sum_{j=1}^m e^{\boldsymbol{z}_j}}, \operatorname{for} i = 1 \ldots k, \boldsymbol{z} \in \mathbb{R}^{k}$, and $\operatorname{norm}(\cdot)$ is layer normalization function. Essentially, the dual-attention mechanism is performed by feeding different input matrices to Eq. (\ref{eq:attn}) for different levels of attention. We illustrate the dual-attention process with $\mathcal{X}^{TC}$ as an example below. We first apply an ROI-wise attention to compute the attention between the ROIs in each subject. That is, the input matrix for the ROI-attention function $\operatorname{Attn}_{ROI}(\cdot)$ is the adjacency matrix of each subject with $k=m$. On top of this ROI-wise attention, we apply subject-wise attention to compute the attention between subjects in each ROI. The input matrix for the subject-wise attention function $\operatorname{Attn}_{subject}(\cdot)$ is the subject-stacked resultant matrix of each ROI, with a dimension of $n^{TC} \times d$, where \jiaxing{$n^{TC}$} is the number of subjects in $\mathcal{X}^{TC}$. Finally, the output of the subject-wise attention is averaged on the subject dimension to obtain the summary feature matrix. The aggregation function $\operatorname{SA}_X(\cdot)$ performed on $\mathcal{X}^{TC}$ can be formally written as:





\begin{equation}
\label{eq:attn_roi}
\boldsymbol{T}_{ROI}^{TC} = \left[ \operatorname{Attn}_{ROI} \left( \boldsymbol{X} \right) : \boldsymbol{X} \in \mathcal{X}^{TC} \right],  
\end{equation}

\begin{equation}
\label{eq:attn_subject}
\boldsymbol{T}_{subject}^{TC} = \left[ \operatorname{Attn}_{subject} \left( \boldsymbol{T}_{ROI}^{TC}(:,i,:) \right) : i = 1, \dots, m \right], 
\end{equation}

\begin{equation}
\label{eq:dual-attn}
\operatorname{SA}_X(\mathcal{X}^{TC}) = \frac{\sum_{j=1}^{n^{TC}} \boldsymbol{T}_{subject}^{TC}(j,:,:)}{n^{TC}},
\end{equation}

\noindent
where $\boldsymbol{T}_{ROI}^{TC}, \boldsymbol{T}_{subject}^{TC} \in \mathbb{R}^{n^{TC} \times m \times m}$ denote two 3-dimensional matrices to store the outputs of the ROI-wise and subject-wise attention functions, and $\left[ \cdot \right]$ denotes the stack operation to combine a set of \jiaxing{2D} matrices to a single \jiaxing{3D} matrix. The subject aggregation function $\operatorname{SA}_A(\cdot)$ performed on the adjacency matrix sets can be defined in a similar way. 

The ROI-wise attention in Eq. (\ref{eq:attn_roi}) aims to extract discriminative-related ROIs and to mitigate the influence of noise (caused by factors such as cardiac/respiratory noise or scanner instability). \jiaxing{We do an empirical study of how ROI-wise attention reduces site-specific noise in Section \ref{sec:case_study}. The goal of subject-wise attention in Eq. (\ref{eq:attn_subject}) is to leverage the node alignment and focus on the most representative subjects in each class. Utilizing ROI- and subject-wise attention to filter crucial information from two different aspects enables us to extract features more effectively from brain networks.} This design of dual-attention can also guarantee the permutation invariance of brain networks, which means no matter how we permute the indices of subjects or ROIs, the output will be the same.
\jiaxing{By averaging over all subjects from the same group to compute the summary graph via Eq. (\ref{eq:dual-attn}), the impact of the number of subjects in different groups is alleviated. This helps mitigate the class imbalance issue in some datasets (e.g., PPMI and ADNI). The contrast between summary graphs can also weaken the sensitivity to the non-representative subjects and thus mitigate overfitting.}

\noindent
\textbf{Pooling with a Contrast Graph.} This module aims to produce a high-quality representation for each input graph by leveraging the group-discriminative information captured in the contrast graph. To achieve a natural and effective utilization of the contrast graph in graph pooling and meanwhile extract high-level node features, we adopt DiffPool \cite{ying2018hierarchical} as the pooling method. The idea is to coarsen the input graph in a hierarchical manner (via layers) such that similar nodes are grouped together into clusters at each layer to extract high-level node representations. Specifically, our ContrastPool takes the input graph with $m$ number of nodes and coarsens it by grouping the input nodes into e.g., $\frac{m}{2}$ number of clusters in a soft manner. The node grouping is performed on an embedded node feature matrix $\boldsymbol{Z}$ and guided by a cluster assignment matrix $\boldsymbol{S}$ that characterizes the node similarity. The output coarsened graph after this pooling layer contains $\frac{m}{2}$ number of nodes, with each node representing a soft cluster of the input nodes. For hierarchical pooling, $L$ pooling layers can be deployed.

In ContrastPool, the cluster assignment matrix can be naturally implemented by the contrast graph as it captures the relatedness of ROIs to the prediction task. \jiaxing{Incorporating the contrast graph into graph pooling as group-based prior knowledge can also alleviate overfitting for datasets with limited scales.} As shown in Eq. (\ref{eq:assign_matrix}), the contrast graph $G_{contrast} = (\boldsymbol{A}_{contrast}, \boldsymbol{X}_{contrast})$ passes through a GNN pooling, $\operatorname{GNN}^{(l)}_{pool}$, to learn a cluster assignment matrix $\boldsymbol{S}^{(l)} \in \mathbb{R}^{m^{(l-1)} \times m^{(l)}}$. Herein, $m^{(l)}$ is a pre-defined number of clusters at layer $l$ controlled by a hyperparameter named the pooling ratio.  

\begin{equation}
\label{eq:assign_matrix}
\boldsymbol{S}^{(l)}=\operatorname{softmax}\left( \operatorname{GNN}_{pool}^{(l)}(\boldsymbol{A}_{contrast}, \boldsymbol{X}_{contrast}) \right).
\end{equation}
To obtain the embedded node feature matrix $\boldsymbol{Z}^{(l)}$ at layer $l$, we pass the output graph $G^{(l-1)}=(\boldsymbol{A}^{(l-1)}, \boldsymbol{H}^{(l-1)})$ from the previous layer through an embedding GNN: 
\begin{equation}
\label{eq:gnn_emb}
\boldsymbol{Z}^{(l)}= \operatorname{GNN}^{(l)}_{emb}(\boldsymbol{A}^{(l-1)}, \boldsymbol{H}^{(l-1)}),
\end{equation}
\noindent
where the initial representation $\boldsymbol{H}^{(0)}$ is the input feature matrix encoded by a GNN, i.e., $\boldsymbol{H}^{(0)} = \operatorname{GNN}_{enc}(\boldsymbol{X})$.

Once we obtain the cluster assignment matrix $\boldsymbol{S}^{(l)}$ and the embedded node feature matrix $\boldsymbol{Z}^{(l)}$, we generate a coarsened adjacency matrix $\boldsymbol{A}^{(l)}$ and a new feature matrix $\boldsymbol{H}^{(l)}$. This coarsening process can reduce the number of nodes to get higher-level node representations. In particular, we apply the following two equations:
\begin{equation}
\label{eq:feat_update}
\boldsymbol{H}^{(l)}=\boldsymbol{S}^{(l)^\mathsf{T}} \boldsymbol{Z}^{(l)} \in \mathbb{R}^{m^{(l)} \times d},
\end{equation}
\begin{equation}
\label{eq:adj_update}
\boldsymbol{A}^{(l)}=\boldsymbol{S}^{(l)^\mathsf{T}} \boldsymbol{A}^{(l-1)} \boldsymbol{S}^{(l)} \in \mathbb{R}^{m^{(l)} \times {m^{(l)}}}.
\end{equation}
Through Eq. (\ref{eq:feat_update} and \ref{eq:adj_update}), ContrastPool is able to produce high-quality representations for each input graph by extracting high-level node representations under the guidance of the contrast graph. \jiaxing{Intuitively, if an edge with two end ROIs $v_1$ and $v_2$ has a high weight in $\boldsymbol{A}_{contrast}$ (i.e., the dual attention of this connection differs significantly in the two groups), the message passing between $v_1$ and $v_2$ will be more pronounced. As a result, the scores of $v_1$ and $v_2$ in $\boldsymbol{S}$ will share more similarity and contribute to the set of clusters in a similar manner. With the help of the entropy loss $\mathcal{L}_{E_1}$ (to be introduced in the subsection of ``Loss Function''), the learning of the assignment scores of a node will be guided to favor only a few clusters. As a result, two ROIs whose connection is highlighted by the contrast graph will be grouped into the same cluster and high-level node features will be extracted.}

\noindent
\textbf{Loss Function.} Technically, an assignment matrix with entries close to uniform distribution could guide the GNN to treat each ROI and subsequently each node cluster produced in hierarchical pooling equally. Thus, besides the commonly-used cross-entropy loss $\mathcal{L}_{cls}$ \cite{cox1958regression} for graph classification, we also adopt the loss in \cite{ying2018hierarchical} to regularize the entropy of the assignment matrix to avoid such equal treatment. Since a fully-connected adjacency matrix of the contrast graph could cause the over-smoothing of GNN, we apply an entropy loss to the adjacency matrix of the contrast graph $\boldsymbol{A}_{contrast}$ to sparsify the matrix. Note that imposing an entropy loss on $\boldsymbol{A}_{contrast}$ is different from thresholding the edges of $\boldsymbol{A}_{contrast}$. The former has an active impact on the learning of $\boldsymbol{A}_{contrast}$ towards the sparse formation, while the latter is a post-process on an already learnt matrix. The two entropy losses shown in Eq. (\ref{eq:entropy_loss_1} and \ref{eq:entropy_loss_2}) can help class-indicative ROIs to stand out and subsequently boost the model performance, as evidenced in our ablation study in Section \ref{sec:ablation}.
\begin{equation}
\mathcal{L}_{E_1}= \frac{1}{L} \sum^L_{l=1} \frac{1}{m^{(l)}} \sum^{m^{(l)}}_{i=1} \operatorname{entropy} \left( \boldsymbol{S}^{(l)}(i,:) \right), 
\label{eq:entropy_loss_1}
\end{equation}

\begin{equation}
\mathcal{L}_{E_2} = \frac{1}{m} \sum^m_{i=1} \operatorname{entropy} \left( \boldsymbol{A}_{contrast}(i,:) \right).
\label{eq:entropy_loss_2}
\end{equation}
The overall loss $\mathcal{L}$ of ContrastPool is defined as Eq. (\ref{eq:total_loss}), where $\lambda_1$ and $\lambda_2$ are trade-off hyperparameters balancing different losses.
\begin{equation}
\mathcal{L} = \mathcal{L}_{cls} + \lambda_1 * \mathcal{L}_{E_1} + \lambda_2 * \mathcal{L}_{E_2}.
\label{eq:total_loss}
\end{equation}

\section{Experimental Study}

In this section, we first introduce the baseline models and the implementation details of our experiments. We then assess the performance of ContrastPool in comparison with the baseline models. We also present three case studies to show how ContrastPool meets fMRI-specific requirements, and meanwhile to provide the domain interpretation of our dual-attention mechanism. We further conduct ablation studies to analyze the effects of the components in our model. Finally, we perform sensitivity tests on the hyperparameters of our model.

\subsection{Baseline Models}
We select various models as baselines, including (1) conventional machine learning models: Logistic Regression, Naïve Bayes, Support Vector Machine Classifier (SVM), k-Nearest Neighbours (kNN) and Random Forest (implemented by the scikit-learn library \cite{pedregosa2011scikit}); (2) general-purposed GNNs: GCN \cite{kipf2016semi}, GraphSAGE \cite{hamilton2017inductive}, GIN \cite{xu2018powerful}, GAT \cite{velivckovic2017graph}  and GatedGCN \cite{bresson2017residual}; (3) \jiaxing{typical graph pooling approaches, DiffPool \cite{ying2018hierarchical}, SAGPool \cite{lee2019self}, HGPSLPool \cite{zhang2019hierarchical} and MEWISPool \cite{nouranizadeh2021maximum};} 
(4) dense contrast graph with SVM: cs \cite{lanciano2020explainable}; and (5) neural networks designed for brain networks: BrainNetCNN \cite{kawahara2017brainnetcnn}, LiNet \cite{li2019graph}, PRGNN \cite{li2020pooling}, \jiaxing{MG2G \cite{xu2021graph},} BrainGNN \cite{li2021braingnn} and BNTF \cite{kan2022brain}. For GNN baseline models, we sparsify all input graphs by keeping 20\% edges with top correlations in $\boldsymbol{A}$, to avoid over-smoothing.

\subsection{Implementation Details}
\label{sec:imple_detail}

In ContrastPool, we adopt a GCN layer for $\operatorname{GNN}_{enc}$, a GraphSAGE layer for $\operatorname{GNN}_{pool}$ and $\operatorname{GNN}_{emb}$, and a sum pooling with a linear layer for the prediction head. For datasets with more than 2 groups (PPMI and ADNI), we use the most extreme groups to construct the contrast graph: CN and SMC vs. LMCI and AD for \jiaxing{the} ADNI dataset; NC vs. PD for PPMI. The settings of our experiments mainly follow those in \cite{dwivedi2020benchmarkgnns}. We split each dataset into 8:1:1 for training, validation and test, respectively. We evaluate each model with the same random seed under 10-fold cross-validation and report the average accuracy.
The hyperparameters are grid searched by Table \ref{tab:hyperparam_setting}.

\begin{table}[h]
\centering
\caption{Hyperparameter settings.}
\begin{tabular}{lc}
\hline
\textbf{batch   size}    & \begin{tabular}[c]{@{}c@{}}4 ( Taowu and Neurocon),  \\  20 (PPMI, ADNI and ABIDE)\end{tabular} \\
\textbf{$\lambda_1$}         & \{10, 1, 0.1\}    \\
\textbf{$\lambda_2$}         & \{1e-2, 1e-3, 1e-4\}  \\
\textbf{pooling   ratio} & \{0.3, 0.4, 0.5, 0.6\}   \\
\textbf{L}               & \{2, 3, 4\}       \\
\textbf{learning   rate} & \{0.02, 0.01, 0.005, 0.001\}   \\
\textbf{dropout}         & \{0, 0.1, 0.2\}    \\ \hline
\end{tabular}
\label{tab:hyperparam_setting}
\end{table}

\begin{table*}[h]
\centering
\caption{Graph Classification Results (Average Accuracy ± Standard Deviation) over 10-fold-CV. The best result is highlighted in \textbf{bold}. The second best result is \underline{underlined}.}
\scalebox{1.0}
{
\begin{tabular}{clccccc}
\hline
\multicolumn{1}{c}{} & Model & Taowu  & PPMI            & Neurocon        & ADNI           & ABIDE          \\ \hline
\multirow{5}{*}{\shortstack{\textit{Conventional}\\ \textit{ML methods}}} & Logistic   Regression & \textbf{77.50}\scriptsize{±7.50}   & 56.50\scriptsize{±11.02}    & 68.50\scriptsize{±15.17} & 61.99\scriptsize{±0.59}   & 65.82\scriptsize{±3.51}   \\
 & Naïve   Bayes         & 65.00\scriptsize{±12.25}   & 58.83\scriptsize{±5.42}    & 63.50\scriptsize{±11.84} & 48.27\scriptsize{±4.44}   & 63.50\scriptsize{±2.69}   \\
 & SVM                   & 65.00\scriptsize{±16.58}   & \underline{63.67}\scriptsize{±5.11}  & \underline{71.50}\scriptsize{±13.93} & 61.77\scriptsize{±0.25}   & 60.67\scriptsize{±3.61}   \\
 & kNN                   & 62.50\scriptsize{±12.50}   & 53.52\scriptsize{±10.34}  & 56.00\scriptsize{±21.77} & 62.59\scriptsize{±1.73}   & 60.37\scriptsize{±5.64}   \\
 & Random   Forest       & 57.50\scriptsize{±22.50}   & 62.23\scriptsize{±4.22}    & 58.50\scriptsize{±11.19} & 61.77\scriptsize{±0.25}   & 61.18\scriptsize{±5.01}   \\ \hline
\multirow{9}{*}{\shortstack{\textit{General-}\\ \textit{purposed}\\ \textit{GNNs}}} & GCN                   & 60.00\scriptsize{±29.15} & 54.02\scriptsize{±9.06}  & 59.00\scriptsize{±20.71}   & 61.57\scriptsize{±0.60} & 60.97\scriptsize{±2.84}   \\
 & GraphSAGE             & 60.00\scriptsize{±33.91} & 55.00\scriptsize{±12.89} & 68.50\scriptsize{±15.17}   & 61.19\scriptsize{±1.72} & 63.09\scriptsize{±3.11}   \\
 & GIN                   & 65.00\scriptsize{±20.00} & 57.90\scriptsize{±8.12}  & 68.50\scriptsize{±15.17}   & 61.87\scriptsize{±0.38} & 57.02\scriptsize{±3.88}   \\
 & GAT                   & 67.50\scriptsize{±22.50}   & 54.98\scriptsize{±8.03}  & 54.00\scriptsize{±15.62} & 61.34\scriptsize{±1.27} & 60.87\scriptsize{±5.02} \\
 & GatedGCN              & 65.00\scriptsize{±22.91} & 52.60\scriptsize{±11.51} & 69.00\scriptsize{±25.48}   & 62.06\scriptsize{±4.53} & 63.60\scriptsize{±4.70} \\
 & DiffPool              & 65.00\scriptsize{±27.84} & 58.00\scriptsize{±11.00} & 62.50\scriptsize{±25.62} & \underline{63.80}\scriptsize{±4.64}   & 63.75\scriptsize{±3.16} \\ 
  & \jiaxing{SAGPool}              & \jiaxing{65.00\scriptsize{±25.50}} & \jiaxing{52.14\scriptsize{±11.69}} & \jiaxing{65.50\scriptsize{±16.95}} & \jiaxing{58.37\scriptsize{±3.89}}   & \jiaxing{63.70\scriptsize{±3.76}} \\
  & \jiaxing{HGPSLPool}              & \jiaxing{60.00\scriptsize{±30.00}} & \jiaxing{52.57\scriptsize{±11.04}} & \jiaxing{68.50\scriptsize{±21.91}} & \jiaxing{59.71\scriptsize{±5.79}}   & \jiaxing{63.60\scriptsize{±3.50}} \\ 
    & \jiaxing{MEWISPool}              & \jiaxing{57.50\scriptsize{±22.50}} & \jiaxing{54.07\scriptsize{±12.15}} & \jiaxing{66.00\scriptsize{±15.94}} & \jiaxing{61.78\scriptsize{±3.10}}   & \jiaxing{63.99\scriptsize{±4.48}} \\ \hline
\multirow{6}{*}{\shortstack{\textit{Models for}\\\textit{brain networks}}} & cs         & \textbf{77.50}\scriptsize{±17.50} & 58.36\scriptsize{±4.40} & 63.50\scriptsize{±11.84} & 62.25\scriptsize{±0.47} & 62.59\scriptsize{±3.14} \\
 & BrainNetCNN           & 65.00\scriptsize{±27.84} & 57.33\scriptsize{±10.32} & 66.00\scriptsize{±22.45} & 61.08\scriptsize{±2.87} & \underline{65.86}\scriptsize{±2.36} \\
 & LiNet                 & 55.00\scriptsize{±24.49} & 60.71\scriptsize{±10.61} & 56.00\scriptsize{±26.91} & 63.64\scriptsize{±1.73} & 58.14\scriptsize{±3.72} \\
 & PRGNN                 & 67.50\scriptsize{±31.72}   & 58.83\scriptsize{±6.89}  & 63.00\scriptsize{±23.37} & 60.71\scriptsize{±2.21} & 60.76\scriptsize{±4.12} \\ 
  & \jiaxing{MG2G}                 &   \jiaxing{57.50\scriptsize{±22.50}}   & \jiaxing{55.45\scriptsize{±10.24}}  & \jiaxing{68.00\scriptsize{±19.77}} & \jiaxing{63.64\scriptsize{±5.10}} & \jiaxing{64.41\scriptsize{±2.16}} \\ 
 & BrainGNN                 & 67.50\scriptsize{±25.12}   & 61.71\scriptsize{±6.05}  & 56.50\scriptsize{±23.03} &  61.05\scriptsize{±1.23} &  62.88\scriptsize{±2.46}\\
  & BNTF                 & 65.00\scriptsize{±21.08}   & 51.60\scriptsize{±6.15}  & 66.00\scriptsize{±11.97} &  61.94\scriptsize{±3.11} &  63.70\scriptsize{±4.84}\\ \hline
 \textit{Ours} & ContrastPool            & \textbf{77.50}\scriptsize{±17.50}   & \textbf{64.00}\scriptsize{±6.63}    & \textbf{75.00}\scriptsize{±15.81}   & \textbf{67.08}\scriptsize{±2.63}   & \textbf{68.63}\scriptsize{±2.65}   \\ \hline
\end{tabular}
}
\label{tab:main_result}
\end{table*}

\begin{table*}[h]
\centering
\caption{Results of more evaluation metrics on Taowu, Neurocon and ABIDE datasets. The best result is highlighted in \textbf{bold}. For multiclass datasets of ADNI and PPMI, all these metrics are the same as accuracy in Table \ref{tab:main_result}.}
\begin{tabular}{l|c|cccc}
\hline
                          & model        & precision     & recall        & micro-F1    & ROC-AUC     \\ \hline
\multirow{2}{*}{Taowu}    & cs           & 76.67 ± 25.09 & \textbf{81.67} ± 14.47 & 79.09 & \textbf{77.50} ± 17.50 \\
                          & ContrastPool & \textbf{78.33} ± 23.64 & 80.00 ± 25.82 & \textbf{79.16} & \textbf{77.50} ± 17.50 \\ \hline
\multirow{2}{*}{Neurocon} & SVM          & 68.33 ± 29.06 & 85.00 ± 32.02 & 75.76 & 65.00 ± 30.00 \\
                          & ContrastPool & \textbf{75.83} ± 18.05 & \textbf{86.67} ± 20.82 & \textbf{79.13} & \textbf{68.33} ± 20.00 \\ \hline
\multirow{2}{*}{ABIDE}    & BrainNetCNN  & 63.79 ± 3.09  & 64.82 ± 4.90  & 64.30 & 66.68 ± 2.53  \\
                          & ContrastPool & \textbf{64.48} ± 4.08  & \textbf{66.10} ± 9.13  & \textbf{65.28} & \textbf{68.16} ± 3.61  \\ \hline
\end{tabular}
\label{tab:other_metric_result}
\end{table*}

The whole network is trained in an end-to-end manner using the Adam optimizer \cite{kingma2014adam}. We use the early stopping criterion, i.e., we stop the training once there is no further improvement on the validation loss during 25 epochs. 
All the codes were implemented using PyTorch \cite{paszke2017automatic} and Deep Graph Library \cite{wang2019deep} packages. All experiments were conducted on a Linux server with an Intel(R) Core(TM) i9-10940X CPU (3.30GHz), a GeForce GTX 3090 GPU, and a 125GB RAM.

\subsection{Comparison with Baselines}

We report the accuracy on 5 brain network datasets in Table \ref{tab:main_result}. Compared with conventional ML methods, general-purposed GNNs and models for brain network do not show a significant advantage \jiaxing{and attain similar performance in most cases. This finding is consistent with the results reported in existing papers \cite{li2021braingnn}.} Our ContrastPool outperforms all \jiaxing{21} GNN and ML baselines on all datasets. In particular, ContrastPool improves over all GNNs specifically designed for brain networks by up to 13.6\%. The average improvement over GNNs is 8.39\%. \jiaxing{The p-values of one-sided paired t-tests comparing our ContratPool with the best model of brain networks on two large datasets, ADNI and ABIDE, are 0.0483 and 0.00798, respectively. This indicates that our model significantly outperforms existing methods on these two datasets. However, on small datasets with large standard deviations in 10 folds, t-tests are highly sensitive to outliers existing in the 10-fold results, thus decreasing the t-statistic calculated and lowering the chance of rejecting the null hypothesis.}

\begin{figure*}[h]
  \centering
    \subfigure[ContrastPool]{
    \includegraphics[scale=0.67]{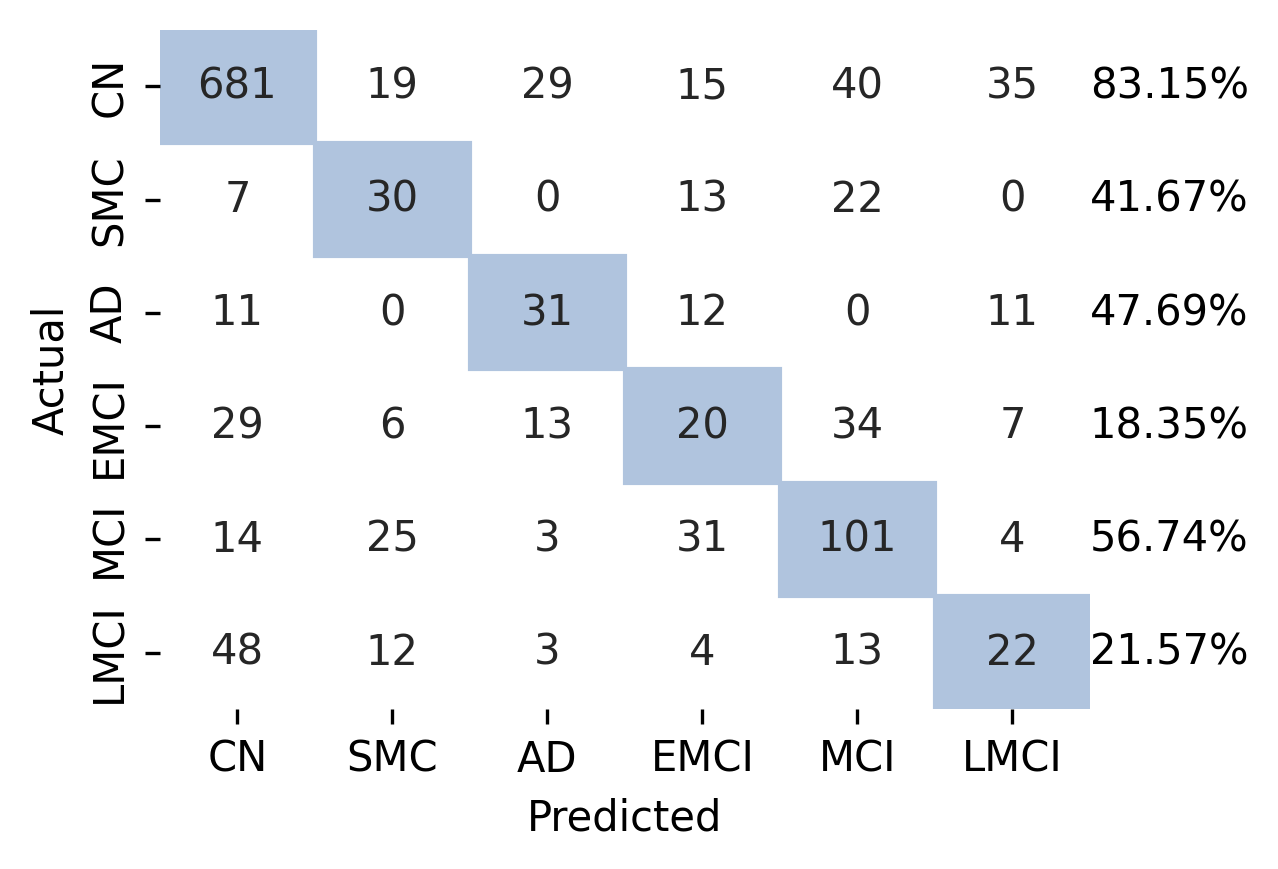}
    }
    \subfigure[BrainGNN]{
    \includegraphics[scale=0.67]{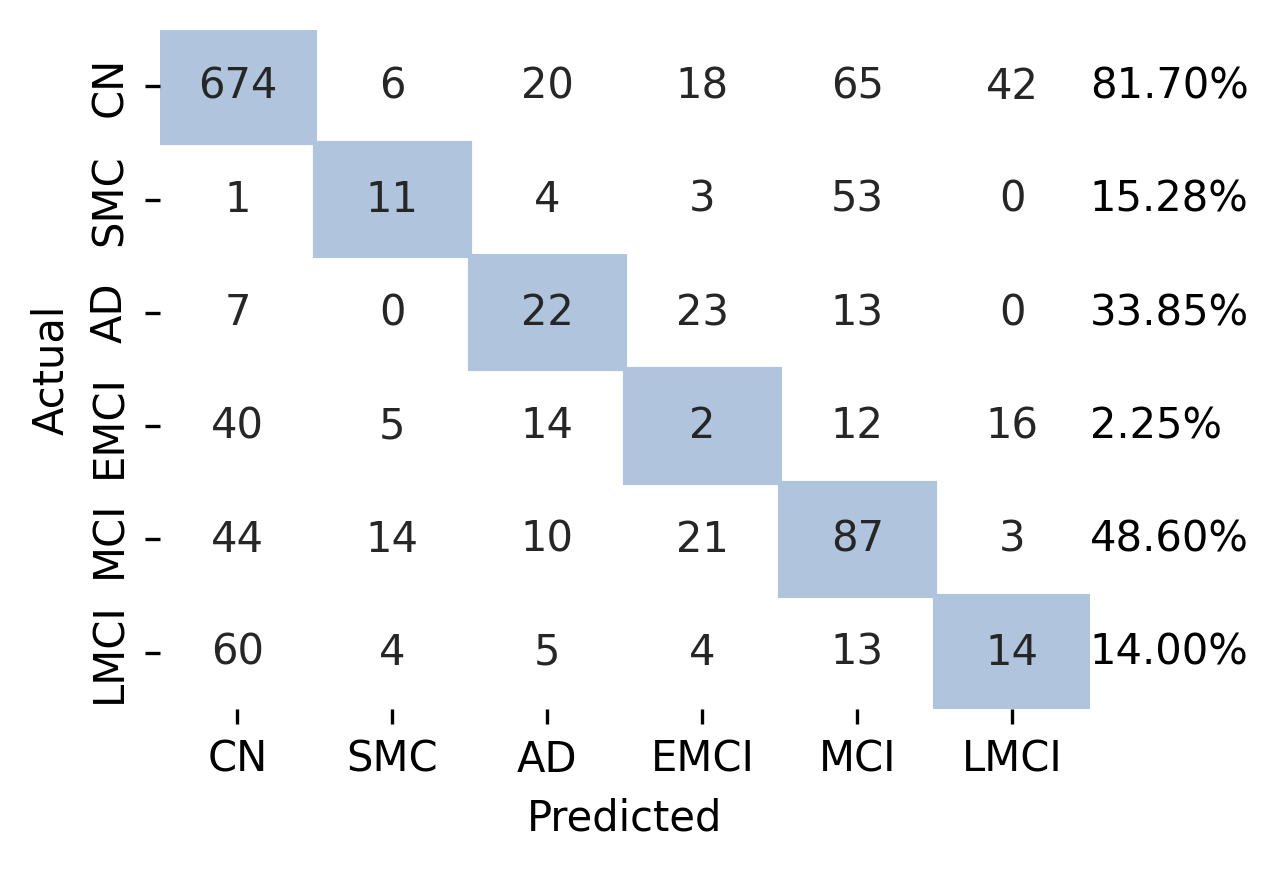}
    }
  \caption{\jiaxing{Confusion matrices with class-wise accuracy of ContrastPool and BrainGNN on the ADNI dataset. Each confusion matrix is obtained by adding the confusion matrices on test sets of all folds.}}
  \label{fig:cm_adni}
\end{figure*}

\jiaxing{It is observed from Table \ref{tab:main_result} that different methods exhibit varied standard deviations on different datasets. Neural network models typically have a larger number of parameters than conventional ML methods, making them susceptible to overfitting, especially on small datasets like Taowu. Consequently, these methods typically exhibit higher standard deviations than conventional ML methods when tested on the same dataset. Regarding the comparison of standard deviation across different datasets, large datasets such as ABIDE are more likely to guarantee consistent data distribution across different folds, and thus tend to achieve smaller standard deviations than small datasets such as Taowu and Neurocon. 
Remarkably, the standard deviation obtained by our ContrastPool lies at the lower end among all with general GNNs and models for brain networks on all datasets. This also demonstrates that ContrastPool is able to remit the problem of overfitting.}

\begin{figure*}[h]
  \centering
    \subfigure[ContrastPool]{
    \includegraphics[scale=0.6]{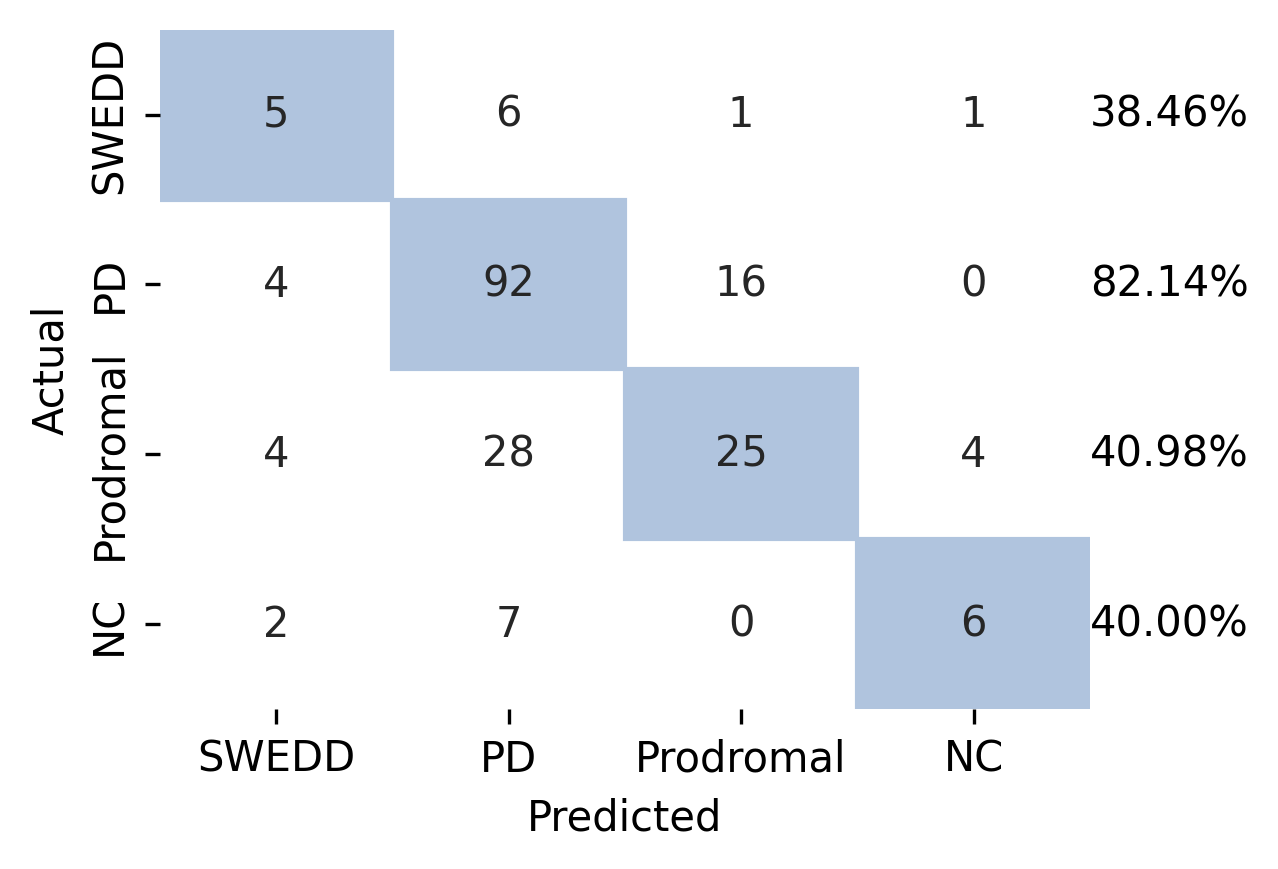}
    }
    \subfigure[BrainGNN]{
    \includegraphics[scale=0.6]{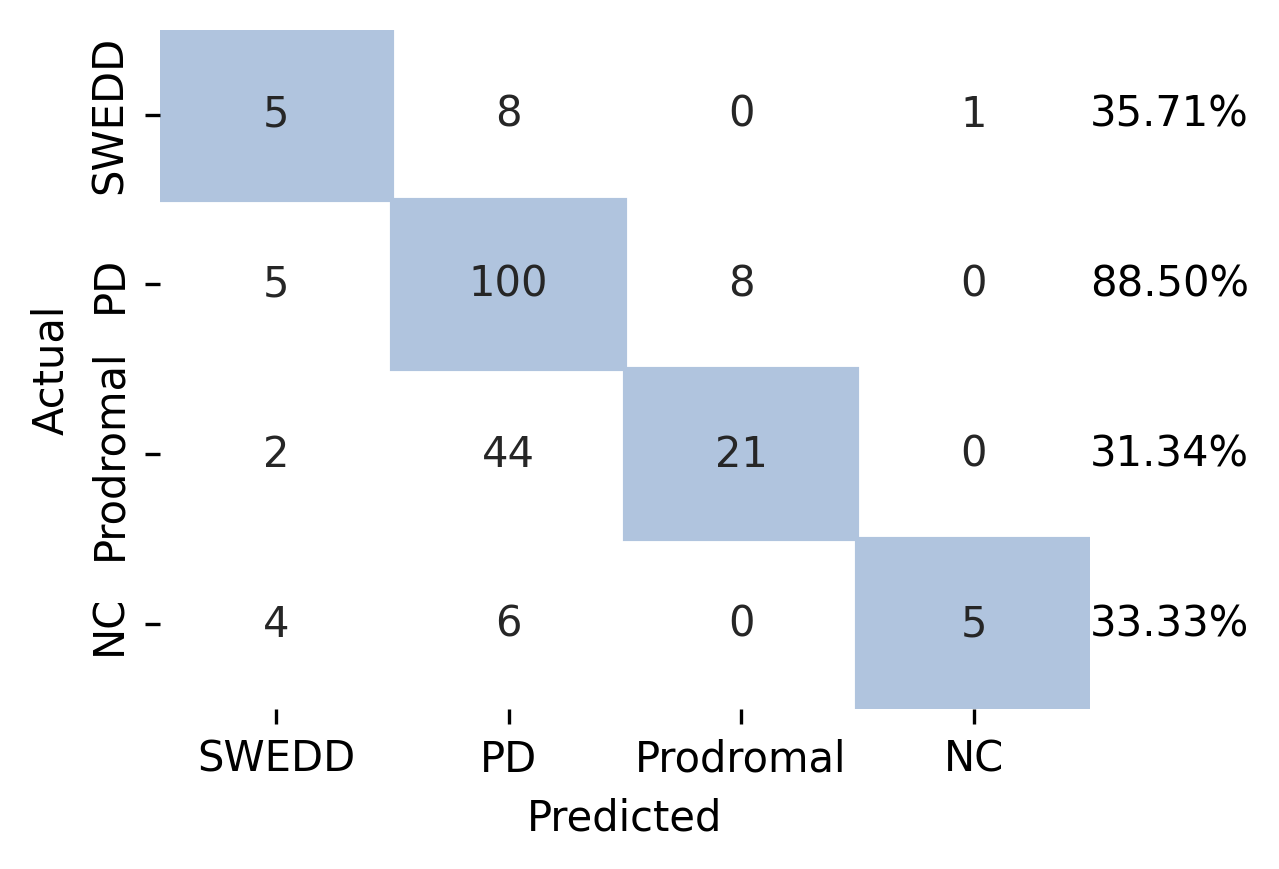}
    }
  \caption{\jiaxing{Confusion matrices with class-wise accuracy of ContrastPool and BrainGNN on the PPMI dataset.}}
  \label{fig:cm_ppmi}
\end{figure*}

\begin{figure*}[h]
  \centering
    \subfigure[\jiaxing{On the ABIDE dataset. Blue edges denote higher attention on TC group and red edges denote higher attention on ASD group.}]{
    \includegraphics[scale=0.3]{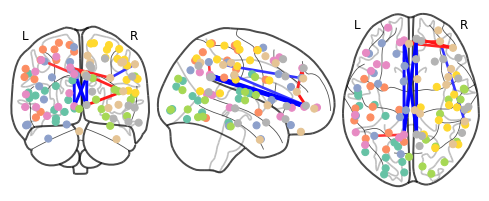}
    }
    \hspace{0.5cm} 
    \subfigure[\jiaxing{On the ADNI dataset. Blue edges denote higher attention on CN/SMC group and red edges denote higher attention on AD/LMCI group.}]{
    \includegraphics[scale=0.3]{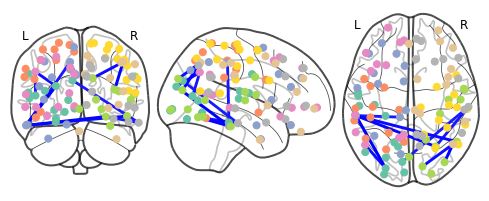}
    }
    \hspace{0.5cm} 
    \subfigure[\jiaxing{On the PPMI dataset. Blue edges denote higher attention on NC group and red edges denote higher attention on PD group.}]{
    \includegraphics[scale=0.3]{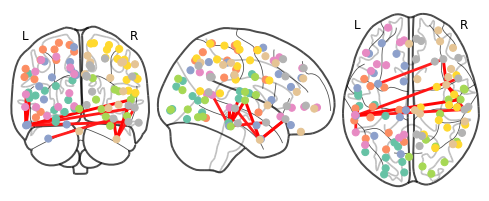}
    } 
  \caption{\jiaxing{Contrast graph visualization. The connections with the top 10 ROI-wise attention weights are highlighted.}}
  \label{fig:contrast_graph}
\end{figure*}

Apart from accuracy, we also report other evaluation metrics, including precision, recall, micro-F1, and ROC-AUC, of the top 2 models on each dataset. As shown in Table \ref{tab:other_metric_result}, ContrastPool performs the best on all datasets over all these metrics except for a single case (recall on Taowu when compared with cs). Note that we do not report the additional metrics on the two multi-class datasets PPMI and ADNI in Table V. This is because in the multi-class case, all these metrics are the same as accuracy, with the superiority of ContrastPool over all baselines already demonstrated in Table \ref{tab:main_result}. 

\jiaxing{To further analyze the performance on multi-class datasets, we plot the confusion matrices with class-wise accuracy on ADNI and PPMI datasets for ContrastPool and BrainGNN in Figs. \ref{fig:cm_adni} and \ref{fig:cm_ppmi}. We can observe that although both ContrastPool and BrainGNN tend to predict subjects as the majority class (CN on ADNI and PD on PPMI), our ContrastPool exhibits higher accuracy on the minority classes (see the column beside each matrix).} 

These experimental results demonstrate the effectiveness of our brain network oriented model design. Existing GNNs with attention mechanism, such as GAT, PRGNN, and BrainGNN, only utilize attention over ROI-level of each subject's feature matrix. To the best of our knowledge, our ContrastPool is the first method that applies ROI- and subject-wise attention on both adjacency matrices and feature matrices. 

\subsection{Case Studies}
\label{sec:case_study}

In this subsection, we present case studies for \jiaxing{ROI-wise attention, subject-wise attention, assignment matrix} and generalization performance to showcase how ContrastPool meets the need of three characteristics of fMRI data.

\begin{figure*}[h]
  \centering
    \subfigure[\jiaxing{On the ABIDE dataset.}]{
    \includegraphics[scale=0.32]{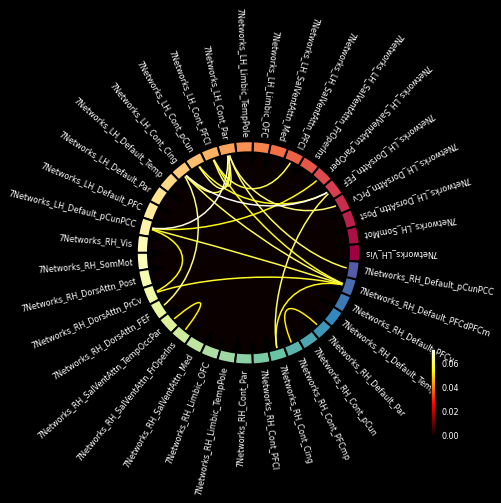}
    }
    \subfigure[\jiaxing{On the ADNI dataset.}]{
    \includegraphics[scale=0.32]{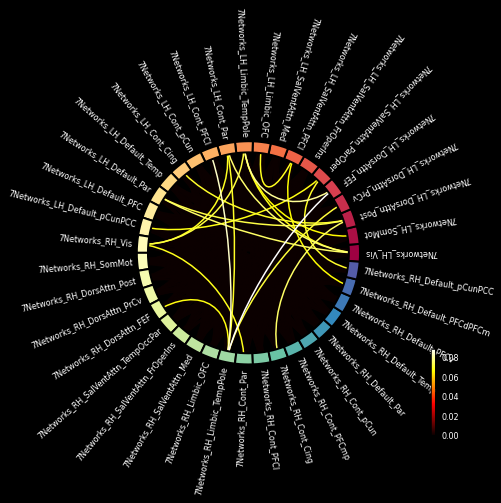}
    }
    \subfigure[\jiaxing{On the PPMI dataset.}]{
    \includegraphics[scale=0.32]{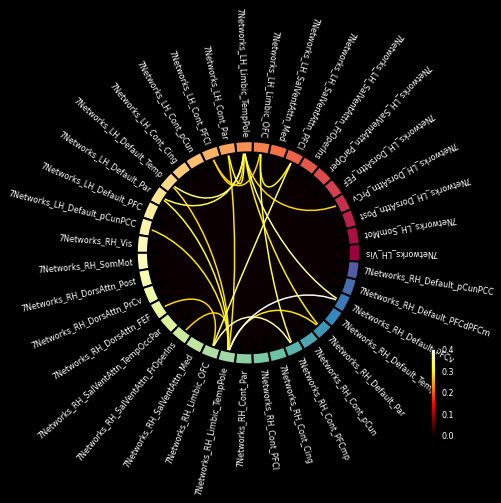}
    } 
  \caption{\jiaxing{Chord diagrams of contrast graphs. Only the edges with top-20 ROI-wise attention scores are shown for better visualization. (a) ROIs related to prefrontal cortex, parietal and cingulate are highlighted for Autism. (b) ROIs related to parietal and posterior are highlighted for Alzheimer's. (c) ROIs related to temporal and ventral prefrontal cortex are highlighted for Parkinson's.}}
  \label{fig:chord_diagram}
\end{figure*}

\noindent
\textbf{Interpretation of ROI-wise Attention.} To interpret the rationality of ContrastPool, we visualize the learnt contrast graphs on ABIDE, ADNI, and PPMI datasets. 
\jiaxing{In order to visualize the ROI-wise attentions that are discriminative across different groups, we plot out the ROI-wise attention scores stored in the contrast graph. Essentially, each ROI attention in the contrast graph is obtained by averaging over all subjects (Eq. (10)) and contrasting between two groups (Eq. (4)).}
As shown in Fig. \ref{fig:contrast_graph}, we select the edges with the top 10 ROI-wise attention weights. \jiaxing{The chord diagram in Fig. \ref{fig:chord_diagram} displays the edges with top-20 ROI-wise attention scores. We merge the ROIs in the same area in the chord diagram for a clearer visualization.} Take the ABIDE dataset in Figs. \ref{fig:contrast_graph}(a) and \ref{fig:chord_diagram}(a) as an example. We can see that the contrast graph differentiates ROI pairs (edges in brain networks) with various importance levels (attention weights) and highlights a number of functional connections with high importance in distinguishing subjects from ASD and TC. Connections between the prefrontal cortex, parietal and cingulate are highlighted by our ROI-wise attention. Some ASD-specific neural machianism \cite{weng2010alterations} may be underlying these connections, and these ROIs were regarded as key involved regions in previous ASD studies \cite{assaf2010abnormal,kana2017neural}. Similar ROI-wise interpretations are found on Alzheimer's and Parkinson's as well. ROIs related to parietal and posterior are highlighted on ADNI (as shown in Figs. \ref{fig:contrast_graph}(b) and \ref{fig:chord_diagram}(b)), while those within temporal and ventral prefrontal cortex are highlighted on PPMI (as shown in Figs. \ref{fig:contrast_graph}(c) and \ref{fig:chord_diagram}(c)). These findings also match the domain knowledge in prior research of Alzheimer's \cite{gould2006task,brier2012loss,alexopoulos2012perfusion,tu2015lost} and Parkinson's \cite{monchi2004neural,gerrits2015compensatory,fernandez2015resting}. 

\jiaxing{We also find some highlighted ROIs that diverge from conventional neuroscientific understanding. For example, posterior has a high attention weight in our model on \jiaxing{the} ADNI dataset. This area belongs to the dorsal attention network, which may imply Alzheimer’s Disease is related to the function of goal-directed, voluntary control of visuospatial attention \cite{fox2006spontaneous}. This insight has not been identified by existing literature \cite{mendez2002posterior}. The attention weights highlighted by ContrastPool may shed some light on neurological pathologies from a data-driven perspective and facilitate related clinical research with unconventional hints.}

\begin{figure}[h]
  \centering
    \subfigure[\begin{scriptsize}\textbf{Site distribution on original features.}\end{scriptsize}]{
    \includegraphics[scale=0.24]{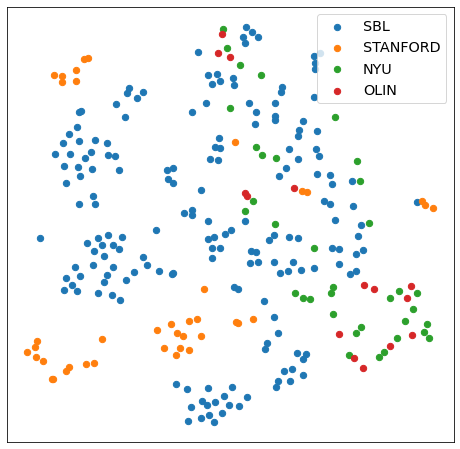}
    }
    \hspace{0.3cm} 
    \subfigure[\begin{scriptsize}\textbf{Site distribution on the representations of ContrastPool w/o subject-wise attention.}\end{scriptsize}]{
    \includegraphics[scale=0.24]{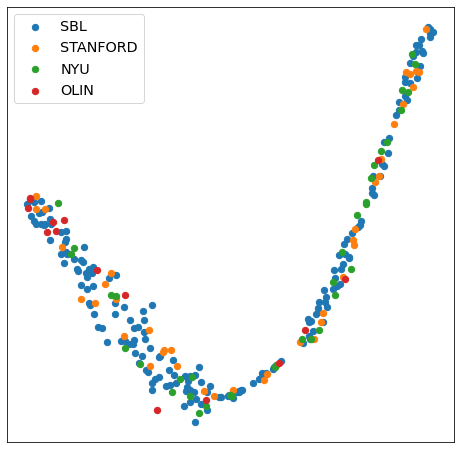}
    } 
  \caption{\jiaxing{The visualization of the original features and ContrastPool representations for subjects from multiple sites on the ABIDE dataset. Each point in the figure represents a subject and different colors denote the sites these subjects are acquired from. The representation of each subject is obtained by mean pooling and visualized by t-SNE\cite{van2008visualizing}. Compared with (b), (a) exhibits obvious site distribution shifts.}}
  \label{fig:feat_dist}
\end{figure}

\jiaxing{We further conducted an experiment to verify the power of the ROI-wise attention in reducing the site-specific noise. We used the ABIDE dataset, which consists of subjects collected from multiple sites and thus may exhibit site differences (scanner variability, different inclusion/exclusion criteria). Fig. \ref{fig:feat_dist} in this document provides a visualization of the site distribution of subjects in the ABIDE dataset under (a) original features and (b) the representations generated by ContrastPool with subject-wise attention disabled. Each point in the figure represents a subject, and different colors denote the sites from which these subjects are acquired. 
Fig. \ref{fig:feat_dist}(a) shows that subjects from different sites do not span over the same subspace but different subspaces. This implies that the site-specific noise may contribute to the distribution shifts under the original features. On the contrary, when we plot out the site distribution under the representations obtained by ContrastPool in Fig. \ref{fig:feat_dist}(b), we observe that such distribution shifts are alleviated: the subjects from different sites span over the same subspace in a more uniform manner. We disable the subject-wise attention so as to focus on the impact from the ROI-wise attention (Eq. (8)). We remark that site information is not utilized at all in the model training and yet ContrastPool is able to reduce such site-specific noise via the design of the ROI-wise attention.}

\begin{figure*}[h]
  \centering
    \subfigure[\jiaxing{On the ADNI dataset. The black and white label denotes AD and MCI subjects, respectively.}]{
    \includegraphics[scale=0.24]{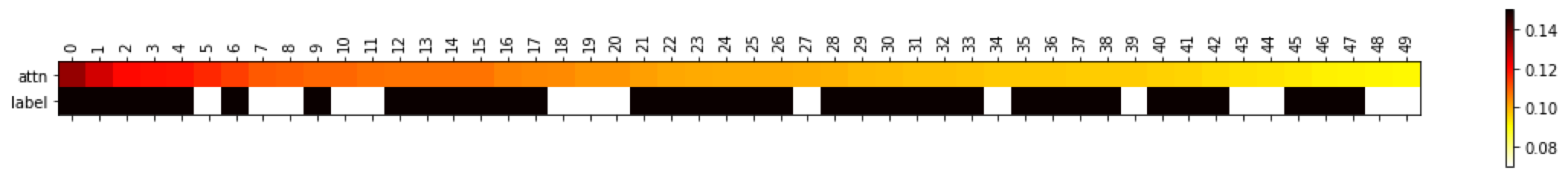}
    }
    \hspace{0.5cm} 
    \subfigure[\jiaxing{On the PPMI dataset. The black and white label denotes PD and SWEDD subjects, respectively.}]{
    \includegraphics[scale=0.22]{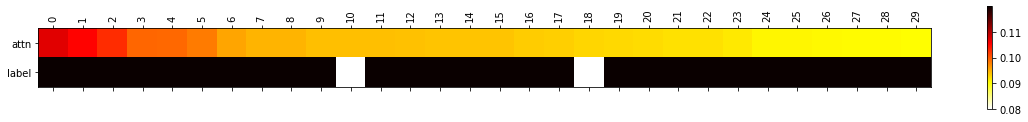}
    }
  \caption{\jiaxing{The heatmap and class distribution on top 50 and top 30 subjects in the subject-wise attention on ADNI and PPMI datasets, respectively.}}
  \label{fig:subject_attn}
\end{figure*}

\noindent
\textbf{Subject-wise Attention.} We design an experiment on \jiaxing{the} ADNI dataset to demonstrate ContrastPool's capability of underscoring informative subjects. Specifically, we merge the MCI
group (82 subjects) with the AD group (143 subjects) into a single group (without revealing to ContrastPool the exact group each subject is from) to be used to contrast against the CN group. 
\jiaxing{Specifically, for a better visualization of the subject-wise attention, we aggregate the weights of all ROIs for each subject into a scalar to represent its subject-wise attention level. Fig. \ref{fig:subject_attn}(a) shows the subjects with the top 50 subject-wise attentions and their class distribution. }
The proportion of AD subjects in the top 50 attention subjects (35/50) is much higher than its proportion in the merged group of subjects (143/225). The top 5 highest attention subjects are all from the AD group. This observation illustrates that subject-wise attention could lead the model to focus more on typical/representative subjects in the dataset.

A similar conclusion can be drawn on PPMI (the other multi-class dataset) as well.  We merge the SWEDD group (
12 subjects) with the PD group (89 subjects) into a single group to be used to contrast against the NC group. Fig. \ref{fig:subject_attn}(b) shows the subjects with the top 30 attention weights. The proportion of PD subjects in the top 30 attention subjects (28/30) is much higher than its proportion in the merged group of subjects (89/101). The top 10 highest attention subjects are all from the PD group, which once again demonstrates the ability of ContrastPool to automatically highlight representative subjects.

\begin{figure}[h]
  \centering
    \includegraphics[scale=0.5]{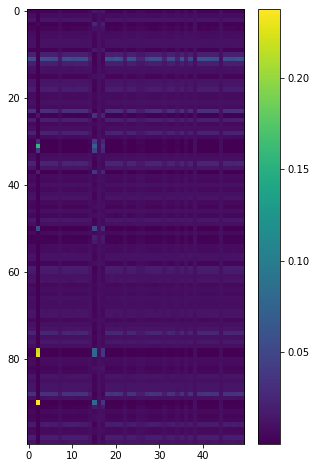}
  \caption{\jiaxing{Visualization of the assignment matrix $\boldsymbol{S}^{(1)}$ at the first layer of ContrastPool on PPMI.}}
  \label{fig:assign_matrix}
\end{figure}

\noindent
\jiaxing{\textbf{Assignment Matrix.} To better understand how the contrast graph works in the differentiable graph pooling, we provide a visualization of the assignment matrix $\boldsymbol{S}^{(1)}$ at the first layer of ContrastPool on the PPMI dataset in Fig. \ref{fig:assign_matrix}. Each row corresponds to an ROI and each column is a cluster. We observe that the differentiable pooling is able to emphasize the most important ROIs and enforce the scores of most of the other ROIs to be close to 0. The entries with high values belong to temporal and ventral prefrontal cortex, which are consistent with the ROIs that have been highlighted in the contrast graph. This demonstrates that the differentiable pooling is well guided by the contrast graph in highlighting the discriminative ROIs.}

\begin{figure}[h]
\begin{center}
\includegraphics[width=\linewidth]{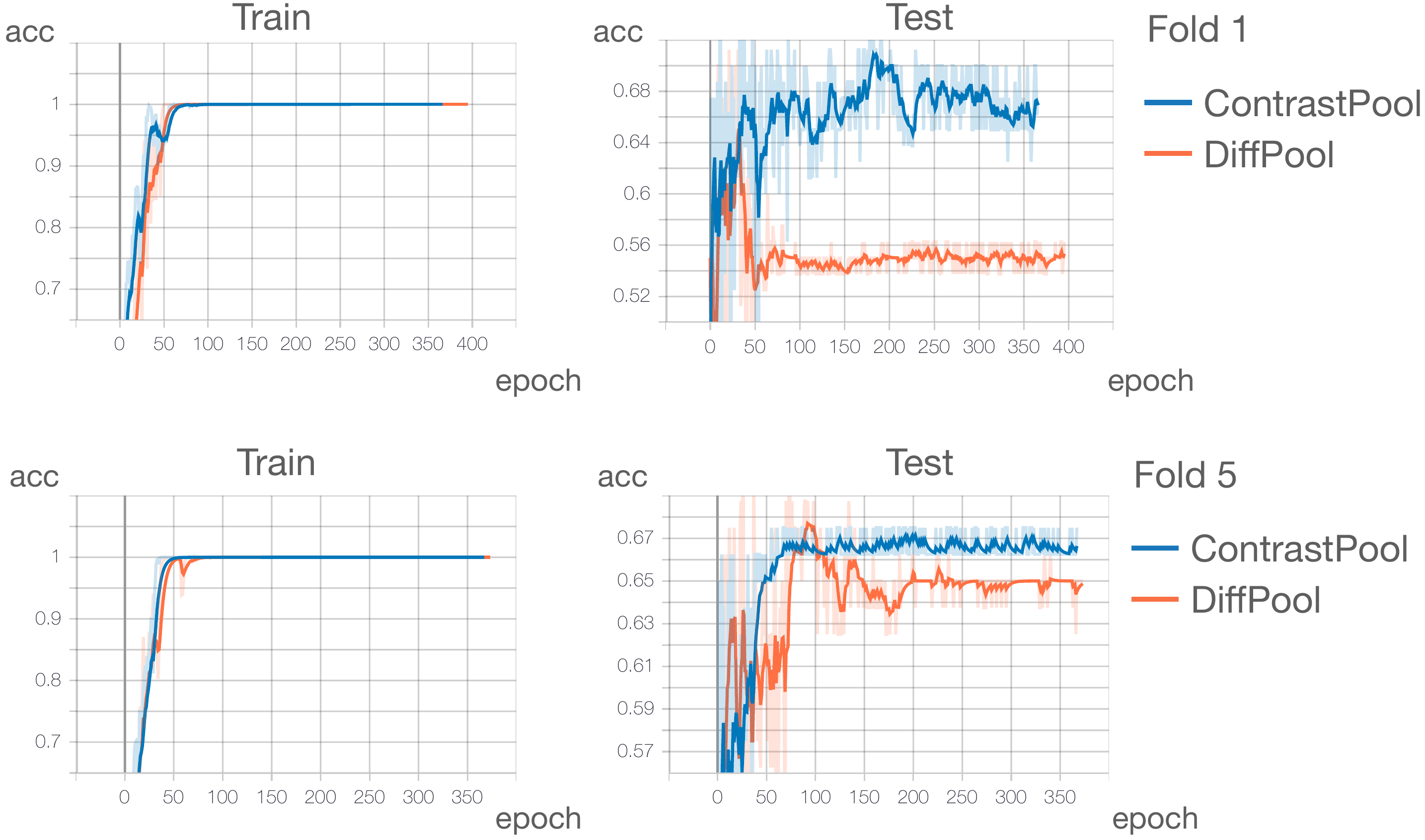}
\end{center}
\caption{Accuracy curves of ContrastPool and DiffPool on two folds on \jiaxing{the} ABIDE dataset.}
\label{fig:acc_curve}
\end{figure}

\noindent
\textbf{Generalization Performance.} We also observe that ContrastPool is able to alleviate the overfitting problem, which is a common problem when applying GNNs to brain networks. An example is shown in Fig. \ref{fig:acc_curve}, where we plot the accuracy curves of ContrastPool and DiffPool on \jiaxing{the} ABIDE dataset. The test accuracy of DiffPool decreases after its training accuracy coverages to 1. ContrastPool \jiaxing{dramatically narrows} the gap between the training and test accuracy, which demonstrates its ability in remitting the overfitting problem.

\subsection{Ablation Studies}
\label{sec:ablation}

In this subsection,  we validate empirically the design choices made in different components of our model: \jiaxing{(1) the dual-attention mechanism and the contrast operation in the Contrastive Dual-Attention (CDA) block; and (2) the loss function. All experiments are conducted on the ABIDE dataset.}

\noindent
\jiaxing{\textbf{Contrastive Dual-attention Block.}} Our CDA block performs dual attention (subject-wise and ROI-wise) on both adjacency matrices and feature matrices. We disable each of them to inspect their effects. The results reported in Table \ref{tab:ablation_cda} demonstrate that the CDA with all \jiaxing{attentions} enabled achieves the best performance. In particular, if the dual attention is entirely disabled (row 1), the summary graph is essentially obtained by averaging over all graphs in each group, which compromises the performance. Moreover, relying solely on ROI-wise attention (row 2) may lead to a further deterioration in performance. This implies that without the support of subject-wise attention, the ROI-wise attention suffers from the information extracted from less representative subjects. 

\jiaxing{To verify the effectiveness of the contrast operation, we conduct an experiment with a CDA variant without considering groups. The results (the last two rows of Table VII) show that ContrastPool with the contrast operation obtains significantly better performance. This verifies the necessity of capturing the discrepancy across groups in the model design.}

\begin{table}[h]
\setlength\tabcolsep{3pt}
\centering
\caption{\jiaxing{Ablation Study on CDA block on ABIDE. Winner is highlighted in \textbf{bold}.}}
\scalebox{1.0}
{
\begin{tabular}{cc|cc|c|c}
\hline
\multicolumn{2}{c|}{\textbf{Adjacency Matrices}} & \multicolumn{2}{c|}{\textbf{Feature Matrices}} & \multirow{2}{*}{\textbf{Contrast}} & \multirow{2}{*}{\textbf{acc ± std}} \\
subject-wise         & ROI-wise         & subject-wise        & ROI-wise        &                      \\ \hline
                     &                  &                     &                & \checkmark  & 65.88 ± 4.11         \\
                     & \checkmark                &                     & \checkmark              & \checkmark  & 64.63 ± 3.83         \\
\checkmark                    &                  & \checkmark                   &             & \checkmark     & 66.25 ± 2.68         \\
                     &                  & \checkmark                   & \checkmark             & \checkmark   & 66.88 ± 4.23         \\
\checkmark                    & \checkmark                &                     &                & \checkmark  & 67.88 ± 4.78         \\
\checkmark                    & \checkmark                & \checkmark                   & \checkmark              &   & 62.71 ± 3.08         \\ 
\checkmark                    & \checkmark                & \checkmark                   & \checkmark              & \checkmark  & \textbf{68.63} ± 2.65         \\ \hline
\end{tabular}
}
\label{tab:ablation_cda}
\vspace{-5pt}
\end{table}


\noindent
\textbf{Loss Function.} We test our design of the loss function by disabling two entropy losses. As shown in Table \ref{tab:ablation_loss}, the results demonstrate that both $\mathcal{L}_{E_1}$ and $\mathcal{L}_{E_2}$ are effective in boosting the model performance.

\begin{table}[h]
    \centering
    \caption{Ablation Study on Entropy Loss on ABIDE. The best result is highlighted in \textbf{bold}.}
    \label{tab:ablation_loss}
    \begin{tabular}{ccc|c}
    \hline
    \textbf{$\mathcal{L}_{cls}$} & \textbf{$\mathcal{L}_{E_1}$} & \textbf{$\mathcal{L}_{E_2}$} & \textbf{acc ± std}          \\ \hline
    \checkmark               &                &       & 64.88 ± 4.45 \\
    \checkmark               & \checkmark              &       & 66.13 ± 2.65 \\
    \checkmark               & \checkmark              & \checkmark     & \textbf{68.63} ± 2.65 \\ \hline
    \end{tabular}
\vspace{-5pt}
\end{table}

\subsection{Hyperparameter Analysis}

In this subsection, we study the sensitivity of three important hyperparameters in ContrastPool, which are the trade-off parameters in Eq. (\ref{eq:total_loss}), the pooling ratio and the number of layers. All experiments are conducted on \jiaxing{the} ABIDE dataset.

\begin{figure}[h]
  \centering
    \includegraphics[scale=0.6]{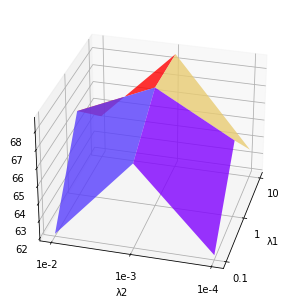}
  \caption{\jiaxing{Results when tuning $\lambda_1$ and $\lambda_2$ on ABIDE.}}
  \label{fig:tune_lambda}
\end{figure}

\noindent
\textbf{Trade-off Parameters $\lambda_1$ and $\lambda_2$.} These two hyperparameters are used in the overall loss function $\mathcal{L}$ (Eq. \ref{eq:total_loss}) for the trade-off between the classification loss and the two entropy losses. We tune the value of $\lambda_1$ from 10 to 0.1 and  $\lambda_2$ from 1e-2 to 1e-4. The results presented in Fig. \ref{fig:tune_lambda} show that our model performs the best when $\lambda_1=1$ and $\lambda_2=$ 1e-3. The same optimal values of $\lambda_1$ and $\lambda_2$ are found on  other datasets.

\begin{figure}[h]
  \centering
    \includegraphics[scale=0.6]{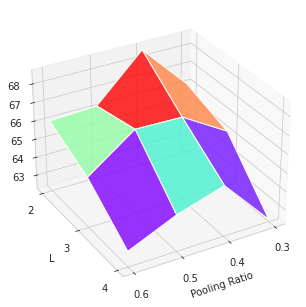}
  \caption{\jiaxing{Results when tuning pooling ratio and the number of layers on ABIDE.}}
  \label{fig:tune_L}
\end{figure}

\noindent
\textbf{Pooling Ratio.} The number of nodes $m^{(l)}$ in the output graph of each ContrastPool layer is controlled by the pooling ratio. A smaller pooling ratio would lead to fewer node clusters in each ContrastPool layer. Herein, we tune it from 0.3 to 0.6. As shown in Fig. \ref{fig:tune_L}, the best choice of pooling ratio is 0.4. The best pooling ratio on different datasets varies slightly within the range of [0.4, 0.5].

\noindent
\textbf{Number of Layers.} The depth of the neural network can undoubtedly affect the model performance. We vary the number of layers $L$ in ContrastPool from 2 to 4, and report the results in Fig. \ref{fig:tune_L}. ContrastPool achieves the best performance when we set $L$ to 2. This indicates that by leveraging the contrast graph, our ContrastPool requires fewer layers to obtain good representations, while most other GNN baselines need to be deeper (e.g., 4 layers) to achieve best performance. The same conclusion of the optimal $L$ can be drawn on other datasets.

\section{Conclusion}
This paper proposes a novel GNN-based solution for brain network classification, taking the unique characteristics of the underlying fMRI data into account. Our proposed method, \textit{ContrastPool}, can adaptively select the most discriminative regions of interest and the most representative subjects by engaging a contrastive dual-attention block. It allows for a flexible local information aggregation within each group. We demonstrate the superiority of our method over \jiaxing{21} state-of-the-art baselines on 5 brain-network datasets spanning over 3 diseases. Moreover, our case studies show the interestingness, simplicity, and high explainability of the patterns extracted by our method, which find consistency in the neuroscience literature. We hope our work can inspire further research in leveraging GNNs for brain network analysis, and show significance in real-world applications, such as the early diagnosis and personalized treatments of neurodegenerative diseases.

\section*{Acknowledgment}

The Neurocon data used in this article were obtained from the NEUROCON project (84/2012), financed by UEFISCDI.

The PPMI data collection used in this article were obtained from the PPMI database (\url{https://www.ppmi- info.org/accessdata-specimens/download-data}). PPMI – a public-private partnership – is funded by The Michael J. Fox Foundation for Parkinson’s Research and funding partners, including 4D Pharma, AbbVie Inc., AcureX Therapeutics, Allergan, Amathus Therapeutics, Aligning Science Across Parkinson’s (ASAP), Avid Radiopharmaceuticals, Bial Biotech, Biogen, BioLegend, Bristol Myers Squibb, Calico Life Sciences LLC, Celgene Corporation, DaCapo Brainscience, Denali Therapeutics, The Edmond J. Safra Foundation, Eli Lilly and Company, GE Healthcare, GlaxoSmithKline, Golub Capital, Handl Therapeutics, Insitro, Janssen Pharmaceuticals, Lundbeck, Merck \& Co., Inc, Meso Scale Diagnostics, LLC, Neurocrine Biosciences, Pfizer Inc., Piramal Imaging, Prevail Therapeutics, F. Hoffmann-La Roche Ltd and its affiliated company Genentech Inc., Sanofi Genzyme, Servier, Takeda Pharmaceutical Company, Teva Neuroscience, Inc., UCB, Vanqua Bio, Verily Life Sciences, Voyager Therapeutics, Inc., Yumanity Therapeutics, Inc.. For up-to-date information on the study, visit \url{www.ppmi-info.org}.

Data collection and sharing of the ADNI data was funded by the Alzheimer's Disease Neuroimaging Initiative (ADNI) (National Institutes of Health Grant U01 AG024904) and DOD ADNI (Department of Defense award number W81XWH-12-2-0012). ADNI is funded by the National Institute on Aging, the National Institute of Biomedical Imaging and Bioengineering, and through generous contributions from the following: AbbVie, Alzheimer’s Association; Alzheimer’s Drug Discovery Foundation; Araclon Biotech; BioClinica, Inc.; Biogen; Bristol-Myers Squibb Company; CereSpir, Inc.; Cogstate; Eisai Inc.; Elan Pharmaceuticals, Inc.; Eli Lilly and Company; EuroImmun; F. Hoffmann-La Roche Ltd and its affiliated company Genentech, Inc.; Fujirebio; GE Healthcare; IXICO Ltd.; Janssen Alzheimer Immunotherapy Research \& Development, LLC.; Johnson \& Johnson Pharmaceutical Research \& Development LLC.; Lumosity; Lundbeck; Merck \& Co., Inc.; Meso Scale Diagnostics, LLC.; NeuroRx Research; Neurotrack Technologies; Novartis Pharmaceuticals Corporation; Pfizer Inc.; Piramal Imaging; Servier; Takeda Pharmaceutical Company; and Transition Therapeutics. The Canadian Institutes of Health Research is providing funds to support ADNI clinical sites in Canada. Private sector contributions are facilitated by the Foundation for the National Institutes of Health (\url{www.fnih.org}). The grantee organization is the Northern California Institute for Research and Education, and the study is coordinated by the Alzheimer’s Therapeutic Research Institute at the University of Southern California. ADNI data are disseminated by the Laboratory for Neuro Imaging at the University of Southern California.

\bibliographystyle{IEEEtrans}
\bibliography{citation}

\end{document}